\documentclass[fleqn,usenatbib]{mnras}
\usepackage[utf8]{inputenc}
\usepackage{newtxtext,newtxmath}

\usepackage[T1]{fontenc}
\DeclareRobustCommand{\VAN}[3]{#2}
\let\VANthebibliography\thebibliography
\def\thebibliography{\DeclareRobustCommand{\VAN}[3]{##3}\VANthebibliography}

\usepackage{graphicx}	
\usepackage{amsmath}	
\usepackage{nicefrac}
\usepackage{orcidlink}

\title[Wind asphericity effects in DF]{The role of wind asphericity in dynamical friction}

\author[J.~D. Carrillo-Santamaria et al.]{
\sloppy
Jes{\'u}s Carrillo-Santamar{\'i}a$^{1}$\thanks{E-mail: jesus.carrillo@correo.nucleares.unam.mx}\orcidlink{0009-0008-0090-7063}, Diego L\'opez-C\'amara$^{1}$\orcidlink{0000-0001-9512-4177},
Fabio De Colle$^{1}$\orcidlink{0000-0002-3137-4633}, Enrique Moreno M\'endez$^{2}$\orcidlink{0000-0002-5411-9352}, \newauthor 
 \ F. J. S\'anchez-Salcedo$^{3}$\orcidlink{0000-0003-2416-2525}
\\
$^{1}$Instituto de Ciencias Nucleares, Universidad Nacional Aut\'onoma de M\'exico, A. P. 70-543 04510 CDMX. Mexico\\
$^{2}$Facultad de Ciencias, Universidad Nacional Aut\'onoma de Mexico, A. P. 70-543 04510 CDMX Mexico\\
$^{3}$Instituto de Astronomía, Universidad Nacional Aut\'onoma de Mexico, A. P. 70-264, 04510. CDMX, Mexico
}

\date{Accepted XXX. Received YYY; in original form ZZZ}

\pubyear{2024}

\begin{document}
\label{firstpage}
\pagerange{\pageref{firstpage}--\pageref{lastpage}}
\maketitle

\begin{abstract}
Dynamical friction (DF) may affect the dynamics of stars moving through dense media. This is the case for stars and compact objects (COs) crossing active galactic nuclei (AGN) discs, stellar clusters, and common envelopes (CE), driving stellar migration.
DF may decelerate the moving stellar object and may also, under certain conditions, produce an acceleration.
In this paper, we study the DF and its effects in the interaction between a star and the ambient gaseous medium through a set of two-dimensional, hydrodynamical numerical simulations using a wind tunnel configuration. Three different stellar wind configurations are considered: isotropic, polar, and equatorial.
We confirm that the DF can decelerate and accelerate the star and find the critical value of the normalized velocity ($u_c$) that marks the transition between these regimes, for the three wind profiles. The value of $u_c$ for the isotropic wind differs slightly from that obtained in the thin shell approximation; for an aspherical wind, it may either be larger or smaller.
Aspherical winds with small $u$ values produce larger accelerations than isotropic winds, while at high $u$ values, they lead to greater deceleration than the isotropic case.
The timescale for DF to substantially affect the velocity of a stellar object is calculated. It is shown to be relevant in AGN discs and CEs.
\end{abstract}


\section{Introduction}
\label{section:intro}
The gravitational coupling between a massive object and the surrounding medium can lead to momentum transfer. In the case of an object moving through a homogeneous gaseous medium, if the object is modeled as a point-like mass (perfect accretor) or as an extended, non-accreting perturber, it always experiences a retarding force \citep{Ostriker_1999, SanchezSalcedo1999, Canto11}. In fact, both the shocked ambient medium (wake) and the accreted gas onto an accretor lead to a drag force (also known as dynamical friction, DF). The morphology of the wake can be significantly modified by magnetic fields \citep[e.g.,][]{SanchezSalcedo2012}, heating feedback \citep[e.g.,][]{masset2017,park2017,li20,Toyouchi2020} or mechanical feedback from outflows \citep{Shima1986,Inaguchi1986,Gruzinov19}. In this paper, we focus on the effect of outflows on the DF experienced by a moving object.

\citet{RephaeliSalpeter1980} estimated the DF when a moving object emits a spherically symmetric wind. They considered the limit in which the ram pressure is sufficiently strong to strip the gas from the outflow. They found the mass outflow results in an increased DF. Their work was analytic, considered only the influence of the wind behind the star, and derived a DF that opposed the motion of the star.  \citet{Shima1986} and \citet{Inaguchi1986} conducted numerical simulations and found that the DF with mass loss is reduced compared to the scenario without mass loss, as the outflow leads to a density enhancement in front of the object and a density reduction behind it. Interestingly, if the wind velocity is supersonic, the body experiences not a DF but a pushing force (negative DF, NDF).

{\citet[][hereafter W96]{Wilkin} derived an analytical solution for the structure of a thin shell bow shock resulting from the interaction between a star — emitting an isotropic wind -  and moving at constant velocity through a uniform-density interstellar medium. \citet[][hereafter G20]{Gruzinov19} computed the DF on a star with an isotropic stellar wind with velocity $v_{w}$, moving with velocity $v_{a}$ through a homogeneous medium, using the analytical solution of W96. They found that the gravitational DF is opposite to its velocity for  $u\equiv v_{a}/v_{w}\gtrsim 1.71$. If $u\lesssim 1.71$, the gravitational force pointed, instead, in the direction of the velocity of the object, that is, NDF. They argued that the effect of NDF was negligible for windy stars moving through the interstellar medium. However, we note that it may be relevant for black holes (BHs) in dense environments because they can drive strong outflows if they accrete mass at a rate well above the Eddington limit. The solution of G20 has been applied to the evolution of a binary system where the two stellar components have isotropic winds \citep{Wang22} and to study the interactions of multiple stars with isotropic outflows inside open clusters \citep{Liu25}.

\citet[][hereafter L20]{li20} investigated, through hydrodynamical simulations, the effect of outflows on the DF experienced by a compact object (CO) in a homogeneous medium. For isotropic outflows with $0.1<u<0.5$,  they found that the strength of the DF was in good agreement with analytical calculations. They also considered accretion-powered jets. In these models, the mass loss rate represents a fraction of the accretion rate. Their simulations demonstrated that the gravitational DF was reduced compared to models without outflows. However, NDF was never achieved for jets aligned with the object's motion, even for $u$ as small as $0.03$. For jets perpendicular to the velocity, NDF was possible but only in highly powerful jets ($u\simeq 0.03$), and only when the mass within several Bondi radii was taken into account. On the other hand, \citet{Toyouchi2020} investigated the effect of radiation on DF, considering a $10^{4}M_{\odot}$ BH embedded in a dusty medium. They found that DF becomes negative if the medium has a density $\lesssim 10^{6}$ cm$^{-3}$ and $v_{a}\lesssim 60$ km s$^{-1}$. \citet{Ogata2021} studied, through numerical simulations, the accretion onto a black hole accretion disc which may be emitting a wind.
DF is relevant in various astrophysical contexts \citep[see][and references therein]{szolgyen2022}. Among the astrophysical objects potentially influenced by DF are: a) Wolf-Rayet stars, which come from massive stars ($M_{\star,ZAMS}\gtrsim20\,M_{\odot}$) that evolve over timescales of a few million years, and exhibit high mass loss rates \citep{WolfRayetbib1,WolfRayetbib2}; 
b) ``immortal stars'', stars that are embedded in extremely dense environments, where the mass gained from the accretion compensates for the wind mass loss, resulting in sustained powerful outflows \citep{Dittmann2021}; d) ``runaway stars'', stellar objects that exhibit anomalously high velocities that have abandoned their birth cluster \citep{runawaystar}; e) young massive clusters (YMCs), which are dense aggregates of young stars ($\sim$100 Myr) formed within galaxies (their mass exceeds $10^4$~$M_{\odot}$ and may have ambient density of $n\sim10^4$~cm$^{-3}$ \citep{starclusterdensity}); and f)  common envelopes (CE), which are an evolutionary phase of close binary systems in which the orbit decays to the point where the secondary star enters the envelope of the primary, subsequently spiraling inward toward the primary's core \citep[e.g.,][]{Chamandy2020}. We investigate whether the DF acceleration can contribute to these phenomena.

In this work, we present a comprehensive study of the DF exerted on a star with a wind, as it moves through a uniform medium. Our study is based on a series of two-dimensional hydrodynamical simulations, in which the effects produced by the DF on the star's velocity are analyzed inside a wind tunnel configuration. In Section~\ref{section:setup} we describe the setup of our simulations. The results are presented in Section~\ref{section:wind_evolution} and ~\ref{section:wind_DF}. We discuss our findings and conclude in Section~\ref{section:astro}.

\section{Setup and units}
\label{section:setup}
\subsection{Numerical setup and models}
\label{section:setup_and_models}
In this work, we aim to determine the amount of DF produced by the interaction between a stellar wind and the environment over which the star moves through. For this purpose, we run a set of two-dimensional (2D) simulations using the hydrodynamical code \textit{Mezcal} \citep{mezcal} in cylindrical coordinates. The adaptive-mesh-refinement code integrates the hydrodynamic equations and is parallelized using the ``Message Passing Interface'' library. 

Figure~\ref{fig1} shows the setups used in our simulations. The star moves at constant velocity through the medium and emits a stellar wind. The coordinate system is set in the co-moving frame of the star (in a ``wind tunnel'' configuration). The star is fixed at the origin of the domain and \textcolor{black}{the ambient medium has a relative velocity $-v_a\hat{\bf z}$}. The ambient medium, an ideal gas with $\gamma = 5/3$, has density $\rho_{a}$, and sound speed $c_s=\sqrt{\gamma P_a/\rho_a}$, where $P_a$ is the ambient pressure. The Mach number of the ambient medium is $M_a=v_a/c_s$.

The stellar wind is launched from the injection radius $r_0$ at constant velocity $v_w$ and constant density $\rho_w$. Thus, the stellar wind has constant mass loss $\dot{M}_w$. The stellar wind can either be isotropic or aspherical. In the aspherical case, we consider two configurations: polar (see left panel of Figure~\ref{fig1}) or equatorial (right panel of the same figure). The opening angle $\theta_w$ is measured from the $Z$-axis for the polar configuration or from the $R$-axis for the equatorial case. 

\begin{figure}	
    \includegraphics[width=\columnwidth]{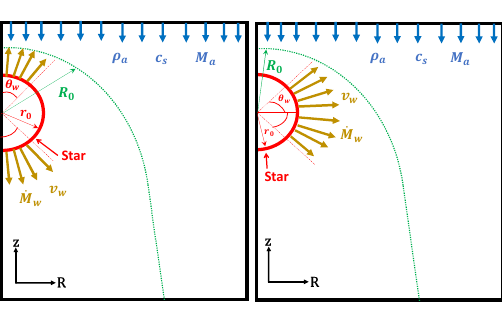}
    \caption{Scheme of the system (not to scale) in a polar (left panel) and equatorial (right panel) configuration. The stellar wind is injected at a distance $r_0$ from the star, with an opening angle $\theta_w$, a mass loss rate $\dot{M}_w$, and a  velocity $v_w$.
    The ambient wind has density $\rho_a$, velocity $v_a$, and sound speed $c_s$. The green line indicates the bow shock.
    $R_0$ is the radius where the stellar and ambient winds balance out.}
    \label{fig1}
\end{figure}

The interaction between the stellar wind and the ambient gas forms a bow shock located (on the $z$ axis) at a distance $R_0$ from the centre of the star. The stellar wind density $\rho_w$ required for the stellar wind ram pressure to balance that of the ambient medium at $R_0$ is:
\begin{equation}
    \rho_w = \frac{\dot{M}_w}{4\pi\left(\cos\theta_1 -\cos\theta_2 \right)r_0^2v_w},
    \label{eq:densidadwind}
\end{equation}
where $\dot{M}_w=4\pi R_0^2\rho_a v_a^2/v_w$ (W96), and $\theta_1 , \theta_2$ are the polar angles limiting the stellar wind injection region, that is: $\theta_{1,2}=(0, \pi/2)$, $(0, \theta_{w})$, $(\pi/2-\theta_{w}, \pi/2)$ for the isotropic, polar, and equatorial cases, respectively. \textcolor{black}{To ensure a stable and smooth transition along the angle $\theta$, we apply an angular smoothing function to the wind density, of the form 
$\rho_w \propto(\theta/\theta_w)^{-10}$ for $\theta\geq\theta_w$, at the injection radius. The angle $\theta$ is measured from the vertical $z$ axis for the polar winds and from the horizontal $R$ axis for the equatorial winds (as shown in Figure~\ref{fig1}).} In terms of $\rho_a$, $v_w$, $R_0$, $r_0$, and $\dot{M}_w$, Equation~(\ref{eq:densidadwind}) can also be written as $\rho_w= (R_0/r_0)^2 \rho_{a} u^2/(\cos\theta_1-\cos\theta_2)$. For the isotropic case $\rho_w=\dot{M}_w/(4\pi r_0^2v_w)$ or, equivalently, $\rho_{w}= (R_0/r_0)^2 u^{2}\rho_{a}$.

At injection, the stellar wind pressure is assumed to be two orders of magnitude lower than its ram pressure: $P_w = \rho_w v_w^2 / 100$. Under these conditions, $P_a \lesssim P_w\ll \rho_{w}v_{w}^{2}$. Therefore, in our simulations, both $P_w$ and $P_a$ are negligible.

We run a set of 2D, hydrodynamic (HD) simulations (31 models) in which we explore a range of values for \textcolor{black}{the velocity ratio between the relative velocity of the ambient medium and the stellar wind velocity ($u=v_a/v_w$)}, varying from $u= 0.07$ to $u=5.0$. For the opening angle, we consider $\theta_w=30^{\circ}, 45^{\circ}$ and $60^{\circ}$, as well as the isotropic case. In all our simulations, the Mach number $M_{a}$ is set to $5$, $c_{s}=1$ and $R_{0}=1$. The total integration time for all models is $t_{f}= 200$. The parameters of each model are given in Table~\ref{table1}.

The computational domain extends from $R_{\rm min}=0$ to $R_{\rm max}=25~R_{0}$ and from $Z_{\rm min}=-40~R_{0}$ to $Z_{\rm max}=10~R_{0}$. The ambient wind is injected at the $Z=Z_{\rm max}$ boundary while the stellar wind is imposed from the injection radius $r_0 = 0.20~R_0$ (which is centred at the origin of the domain). All other boundaries are set with outflow boundary conditions. 
The coarsest level of refinement has ($250\times500$) cells along the $(R,Z)$, corresponding to a maximum resolution of $3\times10^{-3} R_0$. \textcolor{black}{Six resolution levels are used as we verified that with a larger resolution level, the relative difference in the DF is at most $\sim5\%$ (at $z\ll-40$, once steady state has been reached, $F_D^{\infty}$).} 
The stellar wind injection region is resolved with $\sim 200$ cells for the anisotropic cases and $\sim 1000$ cells for the isotropic case. \textcolor{black}{For smaller inner boundaries ($r_0=0.10~R_0$ and $0.15~R_0$), the relative difference in $F_D^{\infty}$ is at most $\sim3\%$ for the isotropic case and {$\sim6\%$} for the equatorial case (both compared to their corresponding case using $r_0=0.20~R_0$). For further details concerning the resolution and the inner boundary, see Appendix~\ref{section:conv}.} 

\begin{table}
\centering
\caption{Model parameters of the 31 simulations. The first column indicates the value of $u=v_a/v_w$, the second column the orientation of the wind: isotropic (Iso), polar (P) or equatorial (E), and the third column the opening angle of the wind $\theta_w$. For the isotropic case, we ran simulations with $u=0.07, 0.2, 0.6, 0.8, 1.3,  1.7, 2.0, 2.6, 3.0, 4.0, 5.0$.}
\begin{tabular}{ccc}
\hline
 $u$ & Orientation &  $\theta_w$ ($^{\circ}$) \\
\hline
 0.07 - 5.0 & Iso  &  -  \\
 0.2 & P, E & 30, 45, 60  \\
 0.6 & P & 45, 60  \\
 2.0 & P & 45, 60  \\
 2.6 & P, E & 45, 60  \\
 5.0 & P, E & 30, 45, 60  \\
\hline
\end{tabular}
\label{table1}
\end{table}

\subsection{Dynamical friction}
\label{section:DF}
The DF will be calculated by \citep{kim}:
\begin{equation}
    F_D =\int \frac{GM_{\star}(\rho-\rho_a)z}{(R^2+z^2)^{3/2}}dV\;,
    \label{eq:intDF}
\end{equation}
where $G$ is the gravitational constant, $M_{\star}$ is the mass of the star, $\rho(R,Z)$ is the density of the environment, shaped by the stellar wind, $\rho_a$ is the ambient density, and $dV= 2 \pi \, R \, dR\, dZ$ is the volume of each cell. 
To calculate the DF at a certain time in our simulations, the $Z$ domain is divided into $350$ equal-sized sections. For each vertical subdivision, the DF is integrated over its total domain $R$. We take the DF to be positive when its direction is parallel to the direction of motion of the star, and negative when it is antiparallel (this is opposed to the reference frame proposed by G20).

The obtained DF is compared to that given by \citet{Ostriker_1999} ($F_D^{Os\,}$), which assumed that the star had no wind, and was given by:
\begin{equation}    
    F_D^{Os} =\frac{4\pi G^2 M_{\star}^2 \rho_a}{v_a^2}\ln\left[ \Lambda \left( 1-\frac{1}{M_a^2} \right)^{\frac{1}{2}} \right] ,
\label{eq:ostriker}
\end{equation}
where $\Lambda=b_{\rm max}/b_{\rm min}$ is the Coulomb impact factor ratio with $b_{\rm max}$ and $b_{\rm min}$ being the maximum and minimum impact parameter values, respectively. For the extension of our computational domain and injection radius, we have $b_{\rm max}=25\,R_{0}$ and $b_{\rm min}=r_0$.

We use dimensionless quantities, which can be converted to real physical units using the appropriate normalization factors. The normalization factors of the variables used in the simulations are the ambient density ($\rho_a'$) for the density, the bow shock radius ($R_0'$) for the length, and the sound speed ($c_s'$) for the velocity; the rest are listed in Table~\ref{table:normalization} and their details are shown in Appendix~\ref{section:norm}.

\begin{table}
\centering
\caption[Rescaling factors.]{Normalization factors (primed values indicate cgs units).}
\begin{tabular}{l l }
\hline
Variable  & Normalization factor \\
\hline
Length    & $R_0'$ \\ 
Density   & $\rho_a'$ \\
Velocity  & $c_s' = v_a'/5$ \\
Mass      & $M'_{\star}$ \\
Pressure  & $\rho_a c_s^2$ \\
Time      & $c_s'/(R'_0 G\ \rho_a')$ \\
Mass transfer rate  & $R_0'^2\rho_a'c_s'u$ \\
DF          & $G\ M_{\star}' \rho_a' R_0'$ \\
Ostriker DF & $G^2 M_{\star}'^2 \ \rho_a' /c_s^2$\\
DF ratio    & $R_0' c_s'^2 / (G M_{\star}')$          \\
\hline
\end{tabular}
\label{table:normalization}
\end{table}

\section{Isotropic and aspherical wind evolution}
\label{section:wind_evolution}
Figure~\ref{fig2} shows the temporal evolution of models with isotropic stellar winds for $u=5.0$ and $0.2$. Specifically, we present density maps and velocity fields at three different times. The first time, $t_1=1$, corresponds to an early stage; $t_i= 20$ corresponds to an intermediate stage; and $t_f =200$ corresponds to the maximum integration time where a stationary state has been reached.  
For $u=5.0$, the wake has high-density and slow material ($\rho \sim 10$, and $v \sim 1$) compared to the ambient ($\rho_a=1$,$v_a=5$). On the contrary, for $u=0.2$ the wake has low-density and fast material ($\rho \sim 10^{-2}$, and $v \sim 20$).
At $t=t_i$, the model with $u=5.0$ presents a more filamentary tail along the $z$-axis compared to $u=0.2$. This occurs because the injected stellar wind velocity is 25 times higher in the $u=0.2$ case. As a result, the stellar material fills the computational domain more rapidly. By $t=t_f$, steady state is reached, and the shocked wind forms an extended and nearly cylindrical structure.
The interaction between the stellar wind and the ambient medium forms a bow shock that encloses a region that contains both shocked ambient material and shocked stellar material. The general structure differs at early and intermediate times and is basically the same and nearly independent of $u$ at $t=t_f$.

In the thin shell approximation, the position of the shock front as a function of the polar angle is given by (W96):
 \begin{equation}
     R(\theta)=\sqrt{3(1-\theta \cot\theta)}\;,
     \label{wilkinteocil}
 \end{equation}
where $\theta=\arccos\left(z/\sqrt{R^2+z^2}\right)$ and $\cot\theta=z/R$. 
This bow shock solution is overlaid in Figure~\ref{fig2} and in Figure~\ref{fig3}.
Our simulations qualitatively reproduce the solution of W96, but show notable differences. While W96 assumed a thin shell with efficient cooling, the latter is neglected in our simulations, resulting in a wide shock structure. 
The position of the bow shock derived from the ram pressure equilibrium between the stellar and ambient winds ($R_0=1$) roughly corresponds in our simulations to the location of the contact discontinuity. However, the bow shock itself is located at $Z(R=0)\sim 1.8$. 

\begin{figure}	
    \includegraphics[width=\columnwidth]{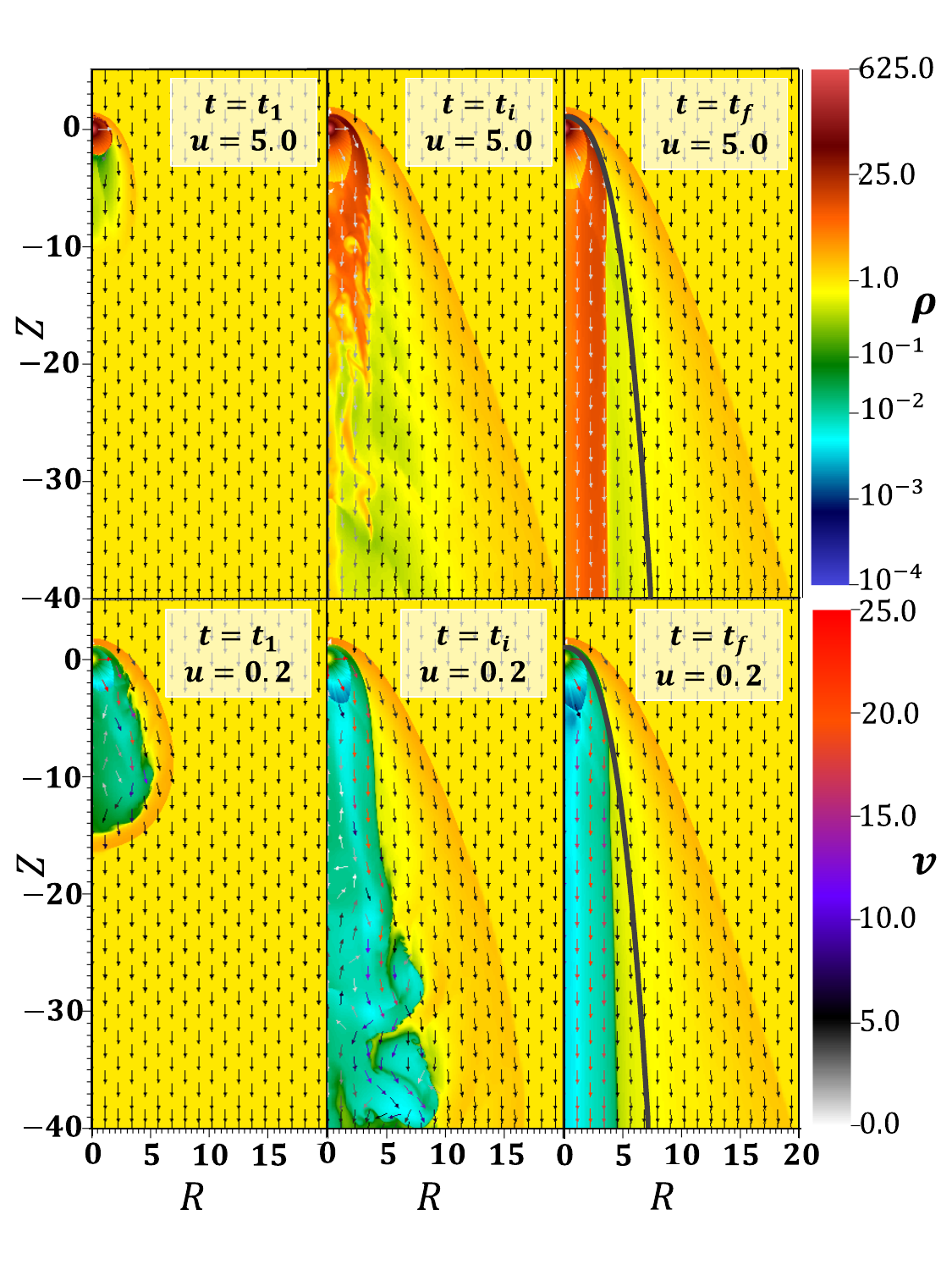}
    \caption{Density maps and velocity fields for isotropic wind models with $u=v_a/v_w=5$ (upper panels) and $u=0.2$ (lower panels). Three different-time snapshots are shown: the initial time ($t_1 = 1$), an intermediate time ($t_i= 20$), and the final time ($t_f=200$), at which a steady state has been reached. The black solid lines represent the analytical bow shock solution from W96.}
    \label{fig2}
\end{figure}

Figure~\ref{fig3} shows that the thin shell approximation does not apply to our models. The simulations reveal four clearly distinct regions: the injected stellar wind, the shocked stellar wind, the shocked ambient medium, and the ambient medium. The shocked stellar wind and the shocked ambient material are separated by a contact discontinuity. This discontinuity is not present in the analytical model of W96.
The density of the ambient medium jumps by a factor $\sim 4$ after the main shock (as expected in the adiabatic case), and the velocity of the unshocked medium and wind change by a factor of $\sim 5$. 
Meanwhile, the shocked stellar wind density depends strongly on the value of $u$.
The normalizations employed imply that $\rho_{w}= 25 \rho_{a} u^{2}$ for the isotropic case (see Section~\ref{section:setup_and_models}).For the $u=0.2$ model, the stellar wind is injected with low density and fast material ($\rho\simeq1$, $v_{w}\approx25$). Meanwhile, the shocked stellar wind has a lower density with faster material ($\rho \sim 10^{-1}$, $v \sim 12$) and the shocked ambient medium has a density of $\rho \sim 4$ and a velocity of $v \sim 4$.
For the model with $u=5.0$, the injected stellar wind is denser and slower ($\rho \sim 625$, $v\lesssim 1$), the shocked stellar wind has a density of $\rho \sim 500$ and a velocity of $v \lesssim 1$, and the shocked ambient medium has a density of $\rho \sim 4$ and a velocity of $v \sim 4$.

\begin{figure}	
    \includegraphics[width=\columnwidth]{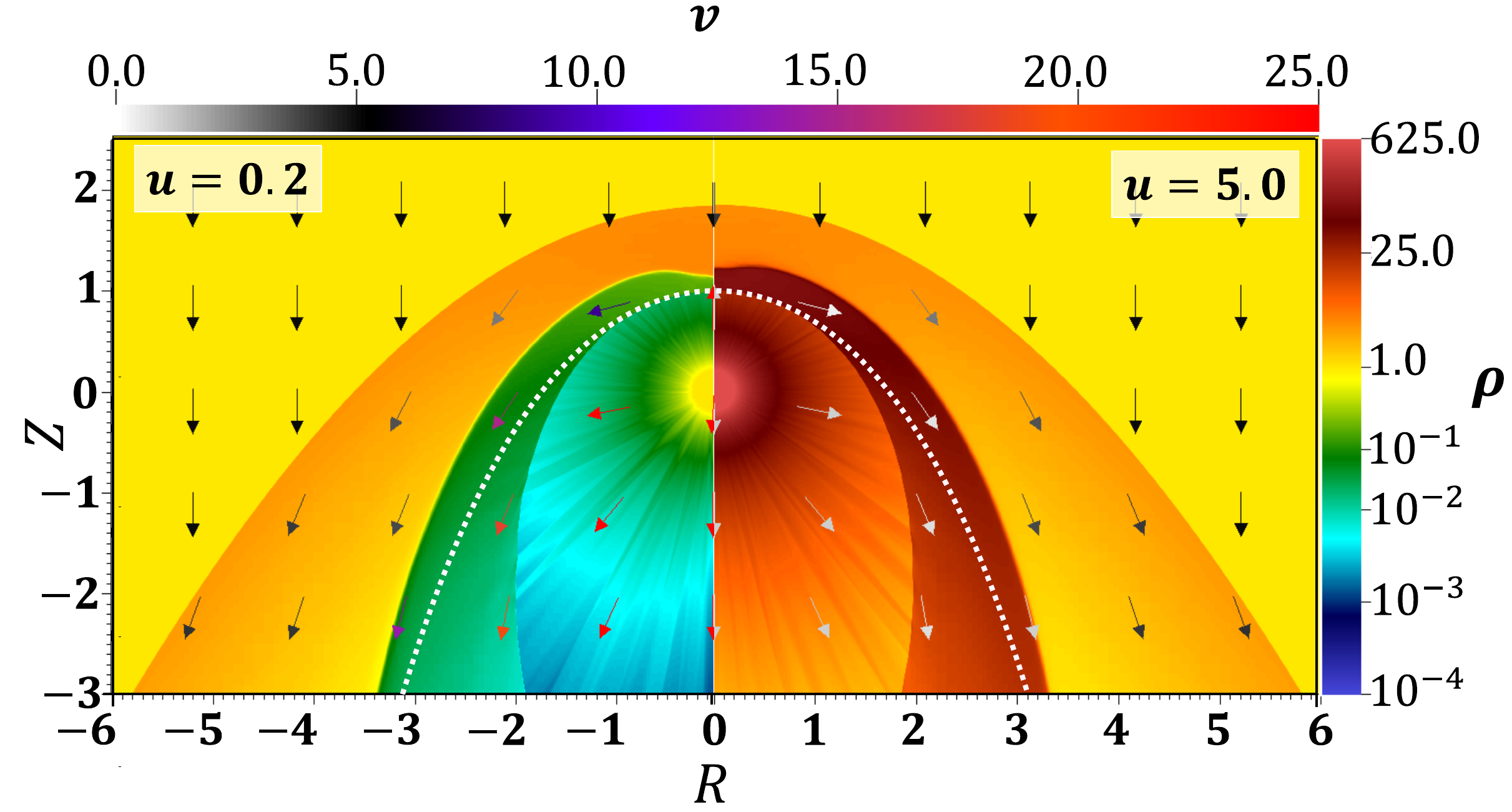}
    \caption{Close-up of the launching region for the isotropic wind models. The $u=0.2$ case (left panel) and the $u=5.0$ case (right panel) are shown. The density, velocity field, and axes are the same as in Figure~\ref{fig1}. The white dotted line represents the analytical bow shock solution from W96. The time shown is $t=t_f$.}
    \label{fig3}
\end{figure}

Figure~\ref{fig4} presents a comparison between simulations with aspherical winds. Specifically, we show density maps and velocity fields for two polar and two equatorial stellar wind models (with $\theta_w = 30^{\circ}$ and $u=5.0$ or $u=0.2$), once they have reached steady state. Wind orientation strongly affects the morphology of the shock. For a polar-oriented wind, the radius at which the ram pressure of the stellar wind balances with that of the ambient medium is $\sim 3$~times farther away from the star compared to the isotropic wind, while for equatorial winds, it is $\sim$~0.7 times farther (also relative to the isotropic wind). 

Regardless of the $u$ value and wind orientation, the shocked ambient medium has densities and velocities similar to those in the isotropic case ($\rho \sim 1$, $v \sim 4$). However, the shocked stellar wind material close to the star (that is, located in the region $-5\leq Z \leq 0$) may reach densities up to one or two orders of magnitude lower than the isotropic case ($\rho \sim 10^{-2}$ and $\rho \sim 10^{-3}$ for the $u=5.0$ and $u=0.2$ cases, respectively).
Far from the star ($-40\leq Z \leq -30$) the shocked stellar wind material may have a higher or lower density than for the isotropic wind depending on the value of $u$. For $u=0.2$ in the equatorial case, the shocked wind has a density an order of magnitude higher than the isotropic case ($\rho \sim 10^{-1}$), and, for the polar case, it has one order of magnitude lower than the isotropic case ($\rho \sim 10^{-3}$); meanwhile, for $u=5.0$ the equatorial shocked wind may have a density that is very similar to that of the isotropical wind ($\rho\sim10$), and the polar shocked wind has a density that is around two orders of magnitude lower than the isotropic case ($\rho \sim 10^{-1}$). Independently of the orientation and the velocity, and as for the isotropic case, the low density material has high velocities and the high density material has low velocities.

\begin{figure}	
    \includegraphics[width=\columnwidth]{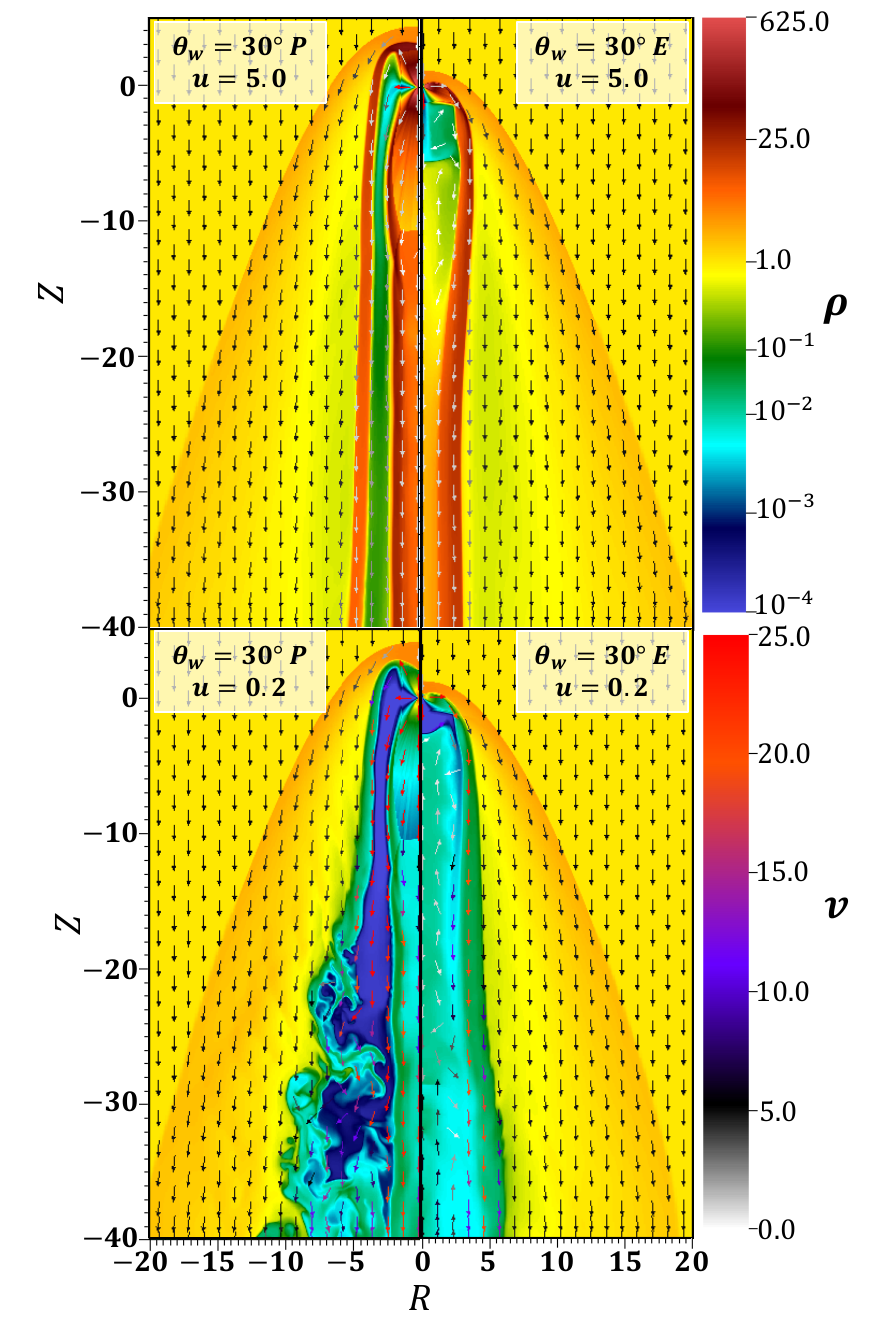}
    \caption{Density maps and velocity fields for aspherical stellar winds at $t=t_f$. Polar winds (left panels) and equatorial winds (right panels) are shown for $u=5.0$ (top) and $u=0.2$ (bottom). In all cases, $\theta_w=30^\circ$. The density, velocity, and axes are the same as in Figure~\ref{fig2}.}
    \label{fig4}
\end{figure}

\section{Dynamical friction effects}
\label{section:wind_DF}
The DF is calculated using Equation~(\ref{eq:intDF}) and the methodology described in Section~\ref{section:DF}. In all figures, the DF is normalized relative to the DF computed by \citet{Ostriker_1999} ($F_D^{Os}$, see Equation~(\ref{eq:ostriker})).

\begin{figure}
    \includegraphics[width=\columnwidth]{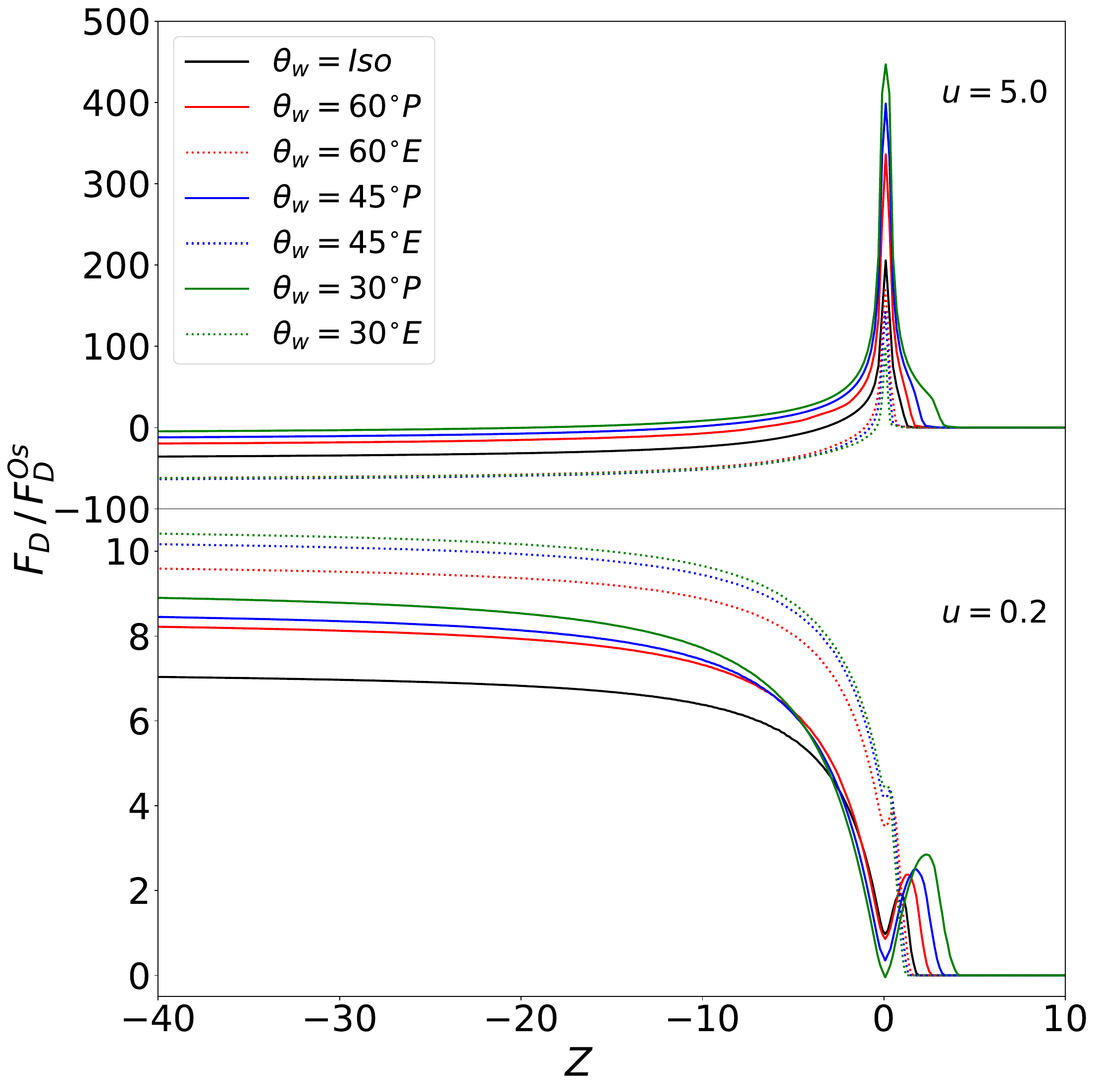}
    \caption{$F_D / F_D^{Os}$ profile for isotropic and aspherical winds as a function of $Z$. The top panel shows models for $u=5.0$ and the bottom panel for $u=0.2$. The black line represents the isotropic case. The green, blue, and red lines indicate $\theta_w= 30^{\circ}, 45^{\circ}, 60^{\circ}$, respectively. Solid lines indicate polar-oriented winds, and dotted lines represent the equatorial ones.}
    \label{fig5}
\end{figure}

\subsection{Distance and velocity dependence}
Figure~\ref{fig5} shows the $F_D / F_D^{Os}$ profiles, integrated from the upper edge of the computational box, up to the distance $Z$ from the star along the symmetry axis (note that the star is located at $R=Z=0$). The figure includes isotropic and aspherical winds (polar and equatorial) with different opening angles and velocity ratios $u=5.0$ and $u=0.2$.

We first describe the behavior of the DF in the spherical case. The contributions to the DF can be divided into two regions: the area around the star and the region well behind it. Regardless of the value of $u$, the shocked region in front of the star has a higher density than the region behind it (see Figure~\ref{fig3}). As a result, the upwind shocked material exerts a positive DF, which tends to accelerate the star. The integrated DF increases and peaks at $Z\approx 0$ (the location of the star) for $u=5.0$ and at $Z \gtrsim 0$ for $u=0.2$. 

Notably, the contribution to the DF from material behind the star strongly depends on $u$, as $u$ dictates the density structure behind the star. For $u=5$, the density in the tail behind the star is higher than that of the unshocked medium, whereas it is lower for $u=0.2$. Consequently, for $u=5.0$ the DF on the star is negative ($F_D / F_D^{Os} \sim -35$), and positive ($F_D / F_D^{Os} \sim 7$) for $u=0.2$, leading to deceleration and acceleration of the star, respectively. The possibility of a pushing force was also discussed by G20, who showed that in the regime where $u\ll 1$, a low-density bubble forms downstream of the star, resulting in a forward-directed force that pushes the star.

Close to the star, at $Z=0$, the behavior of the $u=5.0$ and $u=0.2$ models differs significantly in the aspherical cases. For $u=5.0$, all the curves reach a maximum, while for $u=0.2$, the curves reach a minimum. The peak preceding the minimum in the $u=0.2$ models at $Z>0$ is attributed to the shocked region located in front of the star, where the density is higher than in the region behind it, resulting in a net positive DF (see Figure~\ref{fig3}).

The aspherical cases exhibit the same general trends as the spherical case, with the following differences. In the case of $u=5$, the peak at $Z=0$ increases for more collimated polar winds (as more material is launched in the polar direction). For polar winds, the DF is closer to zero ($F_D / F_D^{Os}$ from $-15 $ to $0$). For equatorial winds, the result is almost independent of the value of $\theta_w$ ($F_D / F_D^{Os} \simeq -80$), suggesting a weaker dependence on the opening angle compared to the polar models. For the $u=0.2$ case, $F_D / F_D^{Os}$ always has positive values, with a minimum at $Z=0$, and increases as the distance from the star increases. The aspherical winds tend to yield larger values with respect to the spherical case. In polar winds, the normalized DF ranges between $8$ and $9$, while equatorial models have the highest DF values, i.e. $F_D / F_D^{Os} \simeq 9.5 - 11$. As the opening angle narrows, the DF increases.

Figure~\ref{fig5} also shows that the DF value converges to a limiting value as $Z$ becomes increasingly negative. To estimate the DF at infinity, that is, $F_D^{\infty} \equiv F_D(Z\rightarrow-\infty)$, we analytically solve Equation~(\ref{eq:intDF}), assuming that the shocked-wind density is constant far from the star and scales as $\propto (R^2+z^2)^{-1}$ close to it. The solution is the following (see Appendix~\ref{ap:solveeq5} for further details):
\begin{eqnarray}
F_D(z) = A \cdot H_1(z) + B \cdot H_2(z) +F_D^{\infty},
\label{eq:fdzinf}
\end{eqnarray}
where $A, B$ are constants, $H_1(z) = (r_0^2+z^2)^{-\frac{1}{2}} - (r_1^2+z^2)^{-\frac{1}{2}}$ and $H_2(z) = (r_0^2+z^2)^{\frac{1}{2}} - (r_1^2+z^2)^{\frac{1}{2}}$. Here, $r_{0}$ is again the injection radius of the wind, and  $r_1$ is the corresponding radius to $F_D^{\infty}$. 
For each model, the $F_D^{\infty}$ value and statistical error were obtained and were always below 4\%. The best fit values of $A, B$ and $r_1$ (using Equation~(\ref{eq:fdzinf})) for representative models (two isotropic wind models and two aspherical wind models) are provided in Table~\ref{tb:constdinf}. The comparison between the fit and the data for the representative models is shown in Figure~\ref{fig:ap1} (isotropic wind with $u=5.0$ and $u=0.2$) and in Figure~\ref{fig:ap2} (polar wind with $u=5.0$ and $\theta_w=45^{\circ}$ and equatorial wind with $u=0.2$ and $\theta_w=60^{\circ}$).

\begin{figure}	
    \includegraphics[width=\columnwidth]{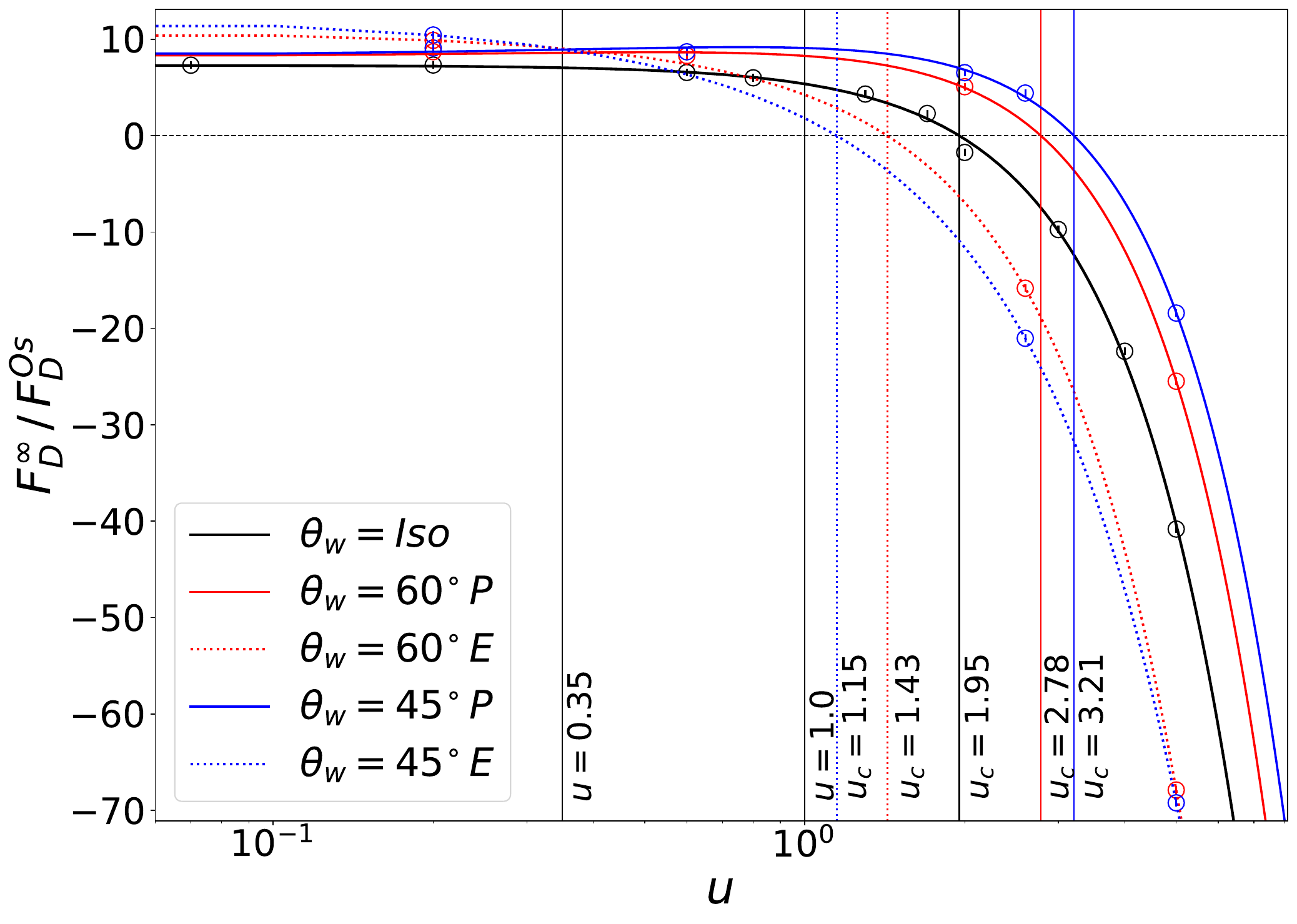}
    \caption{$F_D^{\infty} / F_D^{Os}$ as a function of $u$ for different wind geometries. The isotropic wind case is indicated by the solid black line, polar-oriented winds are indicated by solid colored lines, and equatorial-oriented winds by dotted lines. $\theta_w=60^{\circ}$ is in red and $\theta_w=45^{\circ}$ in blue. The vertical lines indicate the values $u=0.35,\ 1.0,\ 1.15,\ 1.43,\ 1.95,\ 2.78,\ 3.21$ for which the DF is null.}
    \label{fig6}
\end{figure}

Figure~\ref{fig6} shows the dependence of $F_D^{\infty} / F_D^{Os}$ for different winds as a function of $u$. Specifically, we show the isotropic wind model and the polar and equatorial oriented wind models with $\theta_w = 45^{\circ}$ and $\theta_w = 60^{\circ}$. The error bars for each model are included in the figure but are smaller than the size of the data points. For each case, the critical value of $u=u_c$ for which $F_D^{\infty}$ is exactly zero (by fitting the data with a parabola) is obtained. For the isotropic wind, we obtain $u_c=1.95$. For the isotropic wind case, G20 found a slightly lower value of $u_c=1.71$ for the same set of parameters that we used. The difference between the critical values is that we do not assume a thin shell approximation. We expect our value to drop slightly as a function of cooling, approaching $u=1.71$ for a nearly isothermal post-shock region, which is the regime well described by the thin shell approximation. 

Figure~\ref{fig6} shows that the asymptotic value of $F_D^{\infty}$ strongly depends on the value of $u$. For $u<1$, the density in the tail behind the star is lower than the unperturbed ambient density, leading to a positive DF, whereas the opposite occurs for $u>1$. As a result, for $u < 1$, the DF is positive and increases with distance from the star. For $u=1$, in particular, the asymptotic DF is $F^{\infty}_D/F_D^{Os}\approx 5.37$. Also, for $1<u<1.95$, the DF remains positive but decreases with the distance from the star. For $u>1.95$, the DF becomes increasingly negative (i.e., it decelerates the star).

For small $u$ values, aspherical winds have larger accelerations than that for an isotropic wind; larger $u$ values, may produce different outcomes depending on the geometry of the wind. For example, for $u\sim1.5$ values, polar winds accelerate while equatorial winds decelerate; for large $u$ values ($u\gtrsim 3$) equatorial winds produce larger decelerations than the polar case. Independently of the wind orientation, narrower aspherical winds have lower critical $u$ values. For the polar (and equatorial) models with $\theta_w=45^{\circ}$ the critical value is $u_c=2.78$ (and $u_c=1.15$) and for $\theta_w=60^{\circ}$ the critical value is $u_c=3.71$ (and $u_c=1.43$).

\subsection{Opening angle dependence}
Figure~\ref{fig7} shows $F_D^{\infty} / F_D^{Os}$ as a function of the opening angle of the wind for polar and equatorial winds (with $u=5.0$ and $u=0.2$). The error bars for each model are included (and are smaller than the data points for the $u=5$ case). Independently of $u$, the DF for polar models follows a linear fit, with the minimum value being that for the isotropic case. For the $u=5.0$ case, the critical opening angle at which the asymptotic DF will be zero is $\theta_{w,c}=10.16^{\circ}$. 

In order to analyze how the angular dependence of the DF for the polar wind varies as a function of $u$, we run an extra set of polar models (with $\theta_w=45^{\circ}$ and $60^{\circ}$) for different velocity values (from $u=0.2$ to $5.0$). In both the $u\gg 1$ and $u\ll 1$ limits, the DF increases as the opening angle narrows. This result is consistent with the findings of L20, who followed the DF produced by a polar jet with $\theta_w=45^{\circ}$ and found that the positive DF would dominate.

Meanwhile, the DF for the equatorial wind follows a quadratic profile. The DF is always negative for $u=5,0$ (with the isotropic case being the maximum) and positive for $u=0.2$ (with the isotropic model being the minimum). For smaller values of $u$, $F_D^{ \infty}$ shifts toward larger positive values, while it shifts to a more negative value for larger $u$'s. Determining the exact value of the new $u_c$ in this case requires numerical calculations. The magnitude of the $F^{\infty}_D$ for the equatorial winds is larger than that for the polar and isotropic cases.

\begin{figure}	
    \includegraphics[width=\columnwidth]{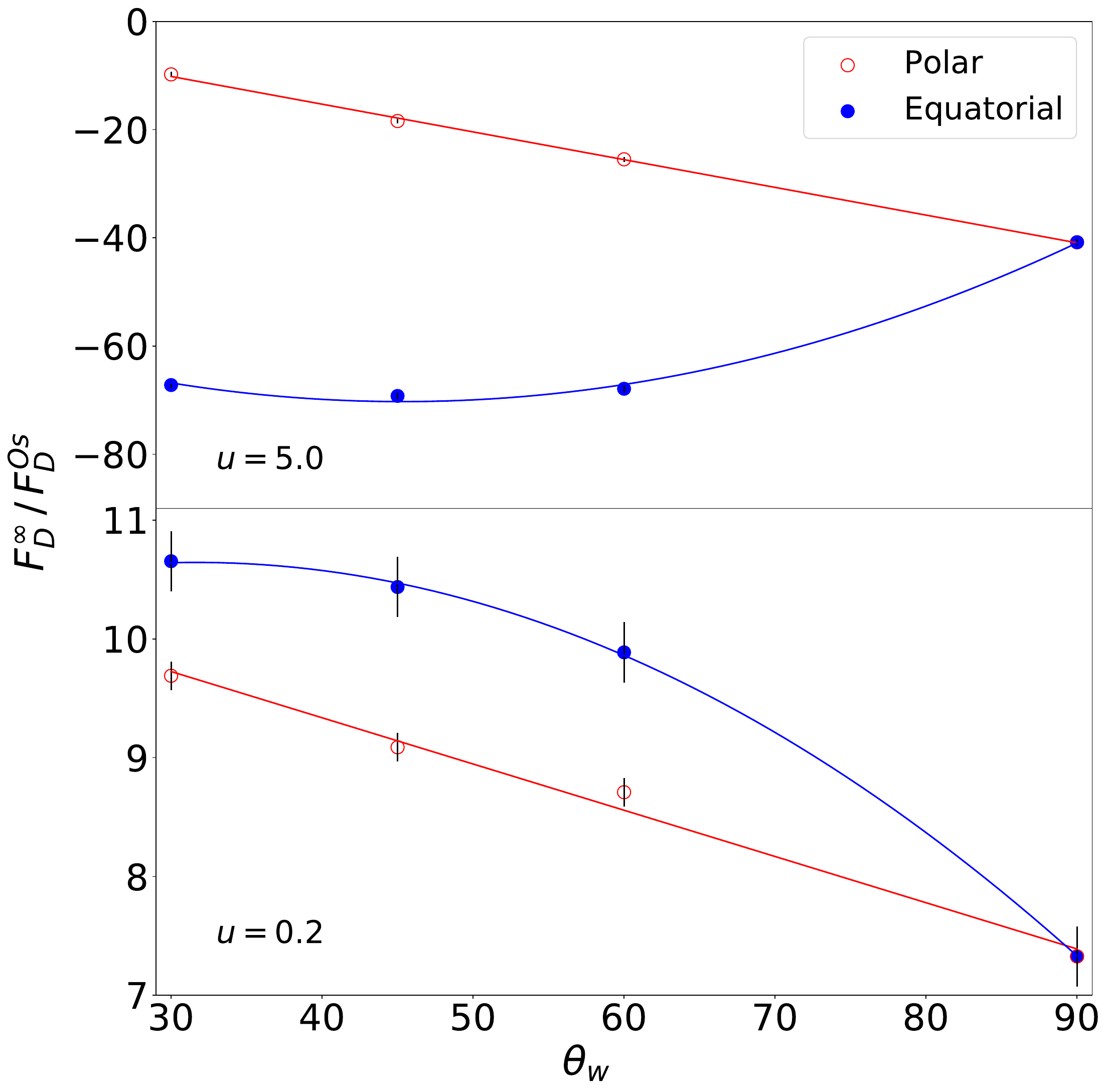}
    \caption{$F_D^{\infty} / F_D^{Os}$ as a function of $\theta_w$ for different winds. The upper and lower panels show the $u=5.0$ and $u=0.2$ cases, respectively. Blue and red dots and lines represent equatorial and polar winds, respectively.}
    \label{fig7}
\end{figure}

\subsection{Timescale}
To obtain $u$ as a function of time, we integrate the normalized DF and its quadratic distribution shown in Figure~\ref{fig6}:
\begin{equation}
    \frac{du}{d\tau} = -\frac{F_D^{\infty}}{F^{Os}_D} = -P_0u^2-P_1u-P_2,
    \label{eq:DFsquare}
\end{equation}
where we have defined the dimensionless time $\tau = t F^{Os}_{D} / (M v_w)$ (see Appendix~\ref{ap:taun} for further details).

For the isotropic wind the best-fit values are $P_0=-1.90$, $P_1=-0.02$, and $P_2=7.29$. Integrating from $u_{\rm min}=u_m$ to $u_{\rm max}=u_c$ for the case where $u<u_c$, and from $u_{\rm min}=u_c$ to $u_{\rm max}=\infty$ for $u>u_c$ we obtain:
\begin{equation}
    u(\tau)=u_m+(u_c-u_m)\cdot\begin{cases}
        \mbox{tanh}\left(\frac{\tau}{\tau_n} \right)\quad\mbox{for}\quad u_0<u_c \\
        \mbox{coth}\left(\frac{\tau}{\tau_n} \right)\quad\mbox{for}\quad u_0>u_c
    \end{cases},
    \label{eq:u(tau)}
\end{equation}
where $u_m\approx-0.03$, $u_c\approx1.95$ and $\tau_n^{-1} = -P_0(u_c-u_m)\sim 4$\footnote{For an isotropic wind $\tau = 0.26$, and for a polar wind with $\theta_w = 45^{\circ}$ or $\theta_w = 60^{\circ}$, $\tau_n=0.27$ and $\tau_n=0.29$, respectively. For an equatorial wind with $\theta_w = 45^{\circ}$ or $\theta_w = 60^{\circ}$, $\tau_n=0.16$ and $\tau_n=0.18$, respectively.}. For further details, see Appendix~\ref{ap:solveeq6}. 
In Figure~\ref{fig8}, we plot $u$ as a function of $\tau / \tau_n$ for an isotropic wind and highlight two regimes ($u<u_c$ and $u>u_c$).

\begin{figure}	
    \includegraphics[width=\columnwidth]{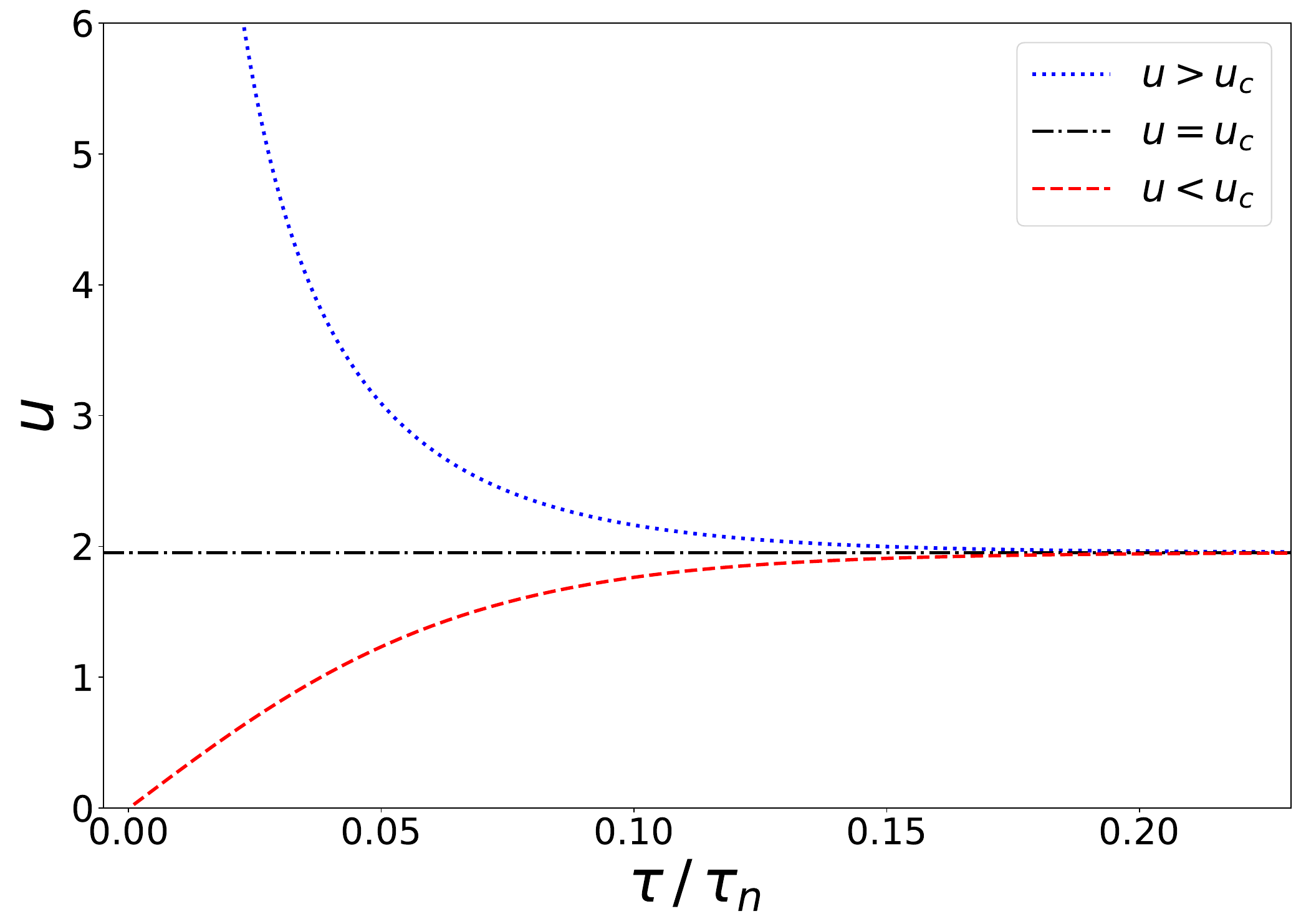}
    \caption{Velocity ratio $u=v_a/v_w$ as a function of time $\tau / \tau_n$ for the isotropic case. The regimes where $u>u_c$ (blue dotted line), $u=u_c$ (black dot-dashed line), and $u<u_c$ (red dashed line) are shown.}
    \label{fig8}
\end{figure}

We estimate the time scale $\tau_c \equiv \tau\left(u=u_c\right)$ required for the star with an isotropic stellar wind to converge to the critical velocity (due to the DF exerted by the ambient medium through which it moves). Regardless of whether the star initially has $u>u_c$ and a negative DF, or $u<u_c$ and a positive DF, its velocity converges to the critical value over a timescale of $\tau_c \sim 0.20\,\tau_n$ (see Figure~\ref{fig7}) for the isotropical and polar oriented winds, and $\tau_c \sim 0.15\,\tau_n$ for the equatorial oriented winds. The critical time in cgs units ($t_c$), for an isotropic wind, is given by (for further details see Appendix~\ref{ap:critime}):
\begin{eqnarray}
    t_c=1.07\times10^{6}\left(\frac{n_a}{ {\rm cm}^{-3}}\right)^{-\frac{1}{2}}\left( \frac{\dot{M}_w}{10^{-14}\,M_{\odot}\,{\rm yr}^{-1}} \right)^{-\frac{1}{2}}\cr \times\left(\frac{v_w}{100\, \rm km\,s^{-1}}\right)^{\frac{1}{2}}\left(\frac{v_a}{100\, \rm km\,s^{-1}}\right)\; {\rm Gyr}\;,
    \label{tauncgs}
\end{eqnarray}
where $n_a$ is the ambient medium number density. This equation adopts typical values for the density of the interstellar medium, and wind velocity, wind mass loss rate, and displacement velocity of Sun-like stars.

\section{Astrophysical applications}
\label{section:astro}
For a star moving through the interstellar medium with $n_a$, $\dot{M}_w$, $v_w$, and $v_a$ equal to the referential values in Equation~(\ref{tauncgs}) we obtain a critical time $t_c\sim 10^5~t_H$, with $t_H$ the Hubble timescale. Also, for G20 (which used $n_a=10$ cm$^{-3}$, $\dot{M}_w=7.24\times10^{-11}\,M_{\odot}$yr$^{-1}$, $v_w=1000$ km s$^{-1}$, and $v_a=10$ km s$^{-1}$), the critical time to reach steady state is well above the Hubble timescale ($t_c \gtrsim 10^3~{\rm Gyr} \gg t_H$). Therefore, the DF is negligible in this case. A higher environment density or a greater stellar mass loss rate is required to reduce the timescale over which the DF becomes important. For example, a Wolf-Rayet star with $\dot{M}\sim (10^{-6}-10^{-4})~M_{\odot}$ yr$^{-1}$ \citep[][]{wolfrayet,wolfrayet2} moving through the same medium will reach the critical velocity in a shorter time ($t_c \sim (1-10)$~Gyr $\lesssim t_H$), but the DF will still remain negligible.

For a Sun-like star moving in an elliptical orbit (hence, with a high Mach number) through the inner part of a disc around an active galactic nucleus (AGN) \citep[specifically, the inner most region of a Seyfert galaxy with $n_a \gtrsim 10^{15}$~cm$^{-3}$, ][]{agndensityJ}, and assuming the same parameters as above, we obtain $t_c \sim 10$~Myr. Thus, the DF is negligible since the timescale is much larger than the translation of the star around a $10^8$~M$_\odot$ supermassive black hole \citep[which is $\sim 90$~days, considering an AGN disc size of 1 lightday,][]{agnlight}. In contrast, the DF acting on a Wolf-Rayet star located within the inner part of an AGN disc is significant (since its critical time is $t_c \sim (100-1000)$~yr) and will cause migration to outer parts of the disc. \textcolor{black}{If the Wolf-Rayet star is located in the outer part of the AGN disc, our results show that it will also experience acceleration due to DF, agreeing with the works of \citep{Liu25}}, but because of the very big critical time $\left(t_c\sim \left(0.4-4\right)\rm{Myr}\right)$, the DF effect is unimportant.

It has been proposed that some of the stars embedded within the external regions of accretion discs of AGNs (with $n_{a}\simeq 10^8$ cm$^{-3}$) could become ``immortal'', as the accretion may counterbalance mass loss \citep{Dittmann2021}. If the outflow winds have velocities equal to a fraction $\lambda$ of the escape velocity $v_{e}$, then $u= M_{a}c_{s}/(\lambda v_{e})$. For $M_{a}=5$, $c_{s}=10$ km s$^{-1}$ and $v_{e}\simeq 10^{3}$ km s$^{-1}$, then $u\leq 0.15$ if $\lambda \geq 1/3$.
Taking $\lambda=1/3$, the outflow is sufficiently powerful to inflate a bubble \citep[i.e. $R_{0}$ is much larger than the Bondi-Hoyle-Lyttleton, BHL, radius,][]{hoylelittleton1939, bondyhoyle1944} if $\dot{M}_{w}\gg 10^{-5}M_{\odot}$ yr$^{-1}$. Thus, only during phases of extremely fast and massive outflows does the immortal star experience a push rather than a pull.

Another relevant case is that of a CO -either a neutron star or a stellar mass black hole (with mass $M_{co}$)- moving through an AGN disc. In this scenario, the CO could accrete at a super-Eddington rate, ejecting part of the accreted material in a collimated outflow. Thus, the outflow mass loss rate is expected to be close to the BHL rate. Assuming $v_a=10$~km~s$^{-1}$ and $v_w\sim c/3$ (this is, $u\sim10^{-4}$), we have:
\begin{equation}
    \dot{M}_{\rm BHL} = \frac{4 \pi \rho_a G^2 M_{co}^2}{v_a^3} \simeq 5.86\times10^{-15}\left(\frac{n_a}{{\rm 1 \; cm}^{-3}}\right) \left(\frac{M_{co}}{ M_\odot}\right)^2 \ M_\odot \;{\rm yr}^{-1}\;.
    \label{bondiacc}
\end{equation}
For the case where the CO is located at the innermost region of the AGN disc ($n_a\sim10^{15}$~cm$^{-3}$), the critical time is $t_c\sim 5f^{-1/2} (M_{co}/M_\odot)^{-1}$~yrs (where $f<1$ is the fraction of the BHL mass accretion rate that feeds the outflow, $\dot{M}_w = f \ \dot{M}_{BHL}$). Thus, the DF is important in the inner AGN disc for a CO. Meanwhile, the DF in the outer part of an AGN disc (where $n_a\sim10^{8}$~cm$^{-3}$) is not of great importance since $t_c\sim 50f^{-1/2} (M_{co}/M_\odot)^{-1}$~Myrs.
Since $u=v_a/v_w \ll 1$, the DF acts by accelerating the CO, causing it to migrate to a wider orbit. As the object moves outwards, the DF decreases, since the timescale for this process scales as $t_c\propto n^{-1} v_a^{-1/2}$. Given that the density in a disc scales as $n\propto r^{-3/2}$ (see G20 and references therein), and that the velocity follows $v\propto r^{-1/2}$ (assuming a keplerian disc), we obtain $t_c\propto r^{7/4}$. This implies that although the DF causes the CO to migrate outwards, its effect gradually drops as the object moves to larger orbital radii.

To explore whether the DF can play a role in accelerating a star out of a YMC and potentially produce a runaway star \citep[for more details, see][]{runawaystar}, we integrate Equation~(\ref{eq:u(tau)}) with respect to time $\tau$ for the case $u_0<u_c$. For Wolf-Rayet stars \citep[with $M\sim 40\,M_{\odot}$ and $v_w=1000$~km~s$^{-1}$ respectively,][]{WolfRayetMass1981} within a massive cluster \citep[with $n_a\sim10^4$~cm$^{-3}$ and a diameter of $\sim20$~pc,][]{20pcCluster}. Assuming that the Wolf-Rayet starts to cross the cluster with a velocity of $v_a=10$~km~s$^{-1}$, the velocity when the end of the cluster is reached is estimated to be  $\sim11.99$~km~s$^{-1}$ (the details of this estimation are shown in Appendix~\ref{ap:rnorm}). This velocity is below the escape velocities of stellar clusters \citep[$\sim17$~km~s$^{-1}$, ][]{Escapevelocity2023}. Thus, the DF-induced acceleration is not a significant mechanism for producing runaway stars.

A CO moving through a CE also accretes at a fraction of the BHL rate. The high accretion rate can lead to the ejection of a jet or outflow with $v_w\gtrsim c/3$ \citep[see, e.g.][]{MorenoMendez2017,LopezCamara2019,LopezCamara2020,LopezCamara2022, Dori2023, Soker2023, Soker2025}. In particular, for a $16~M_{\odot}$~Red Giant (RG) star with a density profile given by $\rho = 0.68 (a/R_{\odot})^{-2.7}$g~cm$^{-3}$ \citep{Papish2015} and an orbital Keplerian velocity $v_a=\sqrt{GM/a} \approx 100 \ (M_{\rm RG}/16 M_\odot)^{1/2} (a/R_\odot)^{-1/2}$ km s$^{-1}$ (where $M_{RG}$ is the mass of the RG), the critical timescale is:
\begin{eqnarray}
    t_c = 60 \ f^{-\frac{1}{2}}\left( \frac{a}{R_\odot} \right)^{2.2} \left(\frac{M_{co}}{ M_\odot}\right)^{-1} \left(\frac{M_{\rm RG}}{16 M_\odot}\right)^{1/2} \; {\rm s}\;.
    \label{tcCE}
\end{eqnarray}
For $f = 0.1$, a $1~M_\odot$ CO, and an orbital separation of $a = R_\odot$, $a = 10~R_{\odot}$ and $a = 100~R_{\odot}$, the steady-state configuration is reached at $t_c\sim 190$~s, $8$~hours and $55$~days respectively. At large orbital separations, the acceleration due to DF is negligible and becomes significant as the CO moves inward. This effect opposes the $\alpha \lambda$ mechanism \citep{vandenHeuvel1976,Eggleton1976,Webbink1984,deKool1987}. The fact that the DF depends on the outflow geometry illustrates the relevance that outflows and jets can have on the dynamical evolution of CEs \citep{Shiber2019}.

\textcolor{black}{Besides the cases of isotropic and polar winds, we note that equatorial winds could be present in a variety of astrophysical scenarios, including massive young stellar objects \citep[YSO, see][]{Hoare2006}, Be stars \citep{decretiondisk2025,Valli25}, or binary systems \citep[i.e. Wolf-Rayet stars,][]{Callingham2019} where the DF and its effects, estimated in this study, may be applied.}

Our results confirm the role of stellar winds in modifying the DF and accelerating the star, as predicted by the analytical estimates of G20. Additionally, we find that this effect also occurs for aspherical winds. A potential caveat is that if the star accretes, the accretion process could suppress the wind ejection, implying that the DF will always act by decelerating the star (see L20 for further details). We argue that this is not necessarily the case, as accretion can suppress wind ejection under the assumptions of Bondi accretion, particularly if the process is spherically symmetric. In a more general scenario of asymmetric accretion, accretion and ejection can occur at the same time, with accretion taking place on the equatorial plane and ejection happening along the polar directions. This  mechanism may operate in many astrophysical phenomena where jets are present, for example: Herbig-Haro \citep{HHjets}, AGN \citep{AGNjets}, tidal disruption events \citep{DeColle2020}, and CEs \citep{Shiber2019, LopezCamara2019}. In the case of a jet launched during the CE phase, \citet{LopezCamara2019} computed this accretion-ejection process self-consistently by assuming that a fraction of the accreted material could power the jet. In this case, the ejection process is expected to be intermittent, fluctuating between phases of high ejection power with low accretion and low ejection power with high accretion.

Our simulations do not include the gravitational force exerted by the star on the gas. This simplification is justified if the BHL radius \citep[$R_{\rm BHL} = 2GM_{*}/v_{a}^{2}$, ][]{hoylelittleton1939, bondyhoyle1944} is  smaller than $\sim 0.25 R_0$ \citep[where $R_0$ is the stand-off radius; for more details see][]{Shima1986}. In terms of $u$, this condition implies that
\begin{equation}
u\lesssim {\frac{\dot{M}_{w}}{60\dot{M}_{\rm BHL}}}\;,
\label{eq:gravity_free}
\end{equation}
where $\dot{M}_{w}$ is the wind mass loss rate of the stellar wind and $\dot{M}_{\rm BHL}$ is the BHL accretion rate. 
If the latter condition is satisfied, the density enhancement in the gravitational wake induced by the star at distances larger than $\sim R_{0}$ can be treated in linear theory. If so, the DF is $F_{D}+F_{D}^{Os}$, where $F_{D}^{Os}$ is given by Equation~(\ref{eq:ostriker}) with $b_{\rm min}\simeq R_{0}$.  The discussed astrophysical cases where the DF plays an important role satisfy this condition. The $u$ value for a Wolf-Rayet star in the inner part of an AGN disc 
is $u \approx 10^{-2}$, while $u \sim 10^{-4}$ for the CO inside a CE. Thus, the condition from Equation~(\ref{eq:gravity_free}) for the Wolf-Rayet case, $u\lesssim0.015$, is satisfied; 
and for the CO inside a CE, $u \lesssim 10^{-3}$, is also satisfied. 

Our simulations have inherent limitations, particularly due to their 2D nature and the chosen coordinate system. Because they are restricted to 2D, they lack turbulence and instabilities which may be present in three-dimensional (3D) simulations. Also, by imposing axis-symmetry, the jet only moves vertically along the polar axis. In contrast, 3D simulations have shown that the jet may wobble around its  axis of motion \citep{LopezCamara2013} and that some degree of asymmetry may be present \citep[e.g.,][]{DuPont_2024}, though this asymmetry is unlikely to significantly impact the DF. Future work will explore 3D models to capture a broader range of jet configurations.  
Other limits are that we neglect curvature effects in the motion of the star, which can be potentially important in the context of CE or AGN discs \citep{kim}. A detailed numerical study of a star or a CO moving through a realistic CE is left for future work.

\textcolor{black}{We caution that assuming a homogeneous medium with no density or velocity gradients may be unrealistic in high-density environments. For example, AGN accretion discs are expected to exhibit a Safronov-Toomre instability criterion \citep{Safronov1960,Toomre1964} parameter near unity \citep[e.g.,][]{Goodman2003,Thompson2005}, implying that they are marginally gravitationally stable. Gravitational fragmentation in such discs could account for in-situ star formation \citep[e.g.,][]{Goodman2003,Nayakshin2006,Levin2007}, suggesting that AGN discs are likely to be clumpy and turbulent. In these self-gravitating, turbulent, and inhomogeneous environments, the torques experienced by embedded objects without winds are found to be stochastic, particularly for those on circular orbits \citep[e.g.,][]{BaruteauPaardekooper, Mayer2013, Malik2015}. If these embedded objects drive winds, the interaction with local substructures is also expected to induce stochastic torques.}

\section{Conclusions}
\label{section:conclusiones}
In this paper, we study through a set of 2D, HD simulations the dynamical friction produced by the interaction between a stellar wind and its environment. We use a ``wind tunnel'' configuration, where the ambient medium has a constant velocity and the star remains fixed. Three different stellar wind configurations are considered: isotropic, polar, and equatorial. Additionally, the wind opening angle and \textcolor{black}{the velocity ratio between the relative velocity of the ambient and the stellar wind velocity ($u=v_a/v_w$) are varied.} The integration time is such that steady state is obtained for all models.

Previous studies focused on how the dynamical friction decelerates a moving stellar object. Recent studies found that acceleration may also be produced by the dynamical friction (see G20 and \textcolor{black}{L20}). We confirm the latter and find the value of $u$ for which each regime takes place in different wind geometries. The critical value of $u$ that separates these two regimes is found for the isotropic wind, and for a subset of polar and equatorial wind models. The $u_c$ obtained for the isotropic wind differs slightly ($\sim 15\%$) from that of G20 where a thin shell was assumed. For the aspherical wind models, the $u_c$ value may be lower or larger ($\sim 65\%$) than that of the isotropic wind.

The interaction between the stellar wind and the ambient medium generates four distinct regions: stellar wind, shocked stellar wind, shocked ambient medium, and ambient medium (the latter two are separated by a bow-shock). For an isotropic wind, the position of the bow-shock is found to differ from the thin shell solution of W96 (nearly double the distance value). 

Aspherical winds produce different dynamical friction values compared to isotropic winds. Aspherical winds with small $u$ values produce larger accelerations than those for an isotropic wind. Meanwhile, aspherical winds with large $u$ values decelerate more than the isotropic case (especially equatorial winds). Narrower winds produce an asymptotic-DF absolute value that is always greater than that for the isotropic case.

The DF effects eventually vanish for every model as the star reaches a critical velocity; then they move at a constant velocity. The acceleration of the star is well described by a quadratic function of the velocity of the star, and the critical time is similar for all wind geometries.

We apply our results to various astrophysical phenomena. We find that the dynamical friction is important for stars and CO in AGN discs and within CEs (and the migration of the star or CO may substantially change). For low density media, like those in YMCs, the DF timescale is so large that its effect is practically negligible compared to other relevant timescales of the system (e.g. the stellar lifetime timescale).

\section*{Acknowledgements}
\label{section:acknowledgements}
JDCS acknowledges support from a SECIHTI fellowship. We acknowledge support from the DGAPA/PAPIIT grant IN113424. We gratefully acknowledge the computing time granted by DGTIC UNAM on the Miztli supercomputer (projects LANCAD-UNAM-DGTIC-281 and LANCAD-UNAM-DGTIC-321).  

\section*{Data Availability}
The data underlying this article will be shared on reasonable request to the corresponding author.

\bibliographystyle{mnras}
\bibliography{bibliografia} 

\appendix
\label{section:appendix}
\section{Convergence}
\label{section:conv}
\textcolor{black}{Figure~\ref{fig:resolution} shows $F_D^{\infty}/F_D^{Os}$ for different levels of refinement ($n=5, 6$, and 7). The model used in all cases is an isotropic wind with $u=5.0$ and the values reported are those obtained once steady state has been reached. We obtain $F_D^{\infty}/F_D^{Os} = -39.11, -40.82, -41.29$ for 5, 6, and 7 levels of refinement, respectively. Thus, the difference in the DF is less than 5\% relative for the resolution we use ($n=6$) compared to a higher resolution level.}
\begin{figure}
    \centering
    \includegraphics[width=\linewidth]{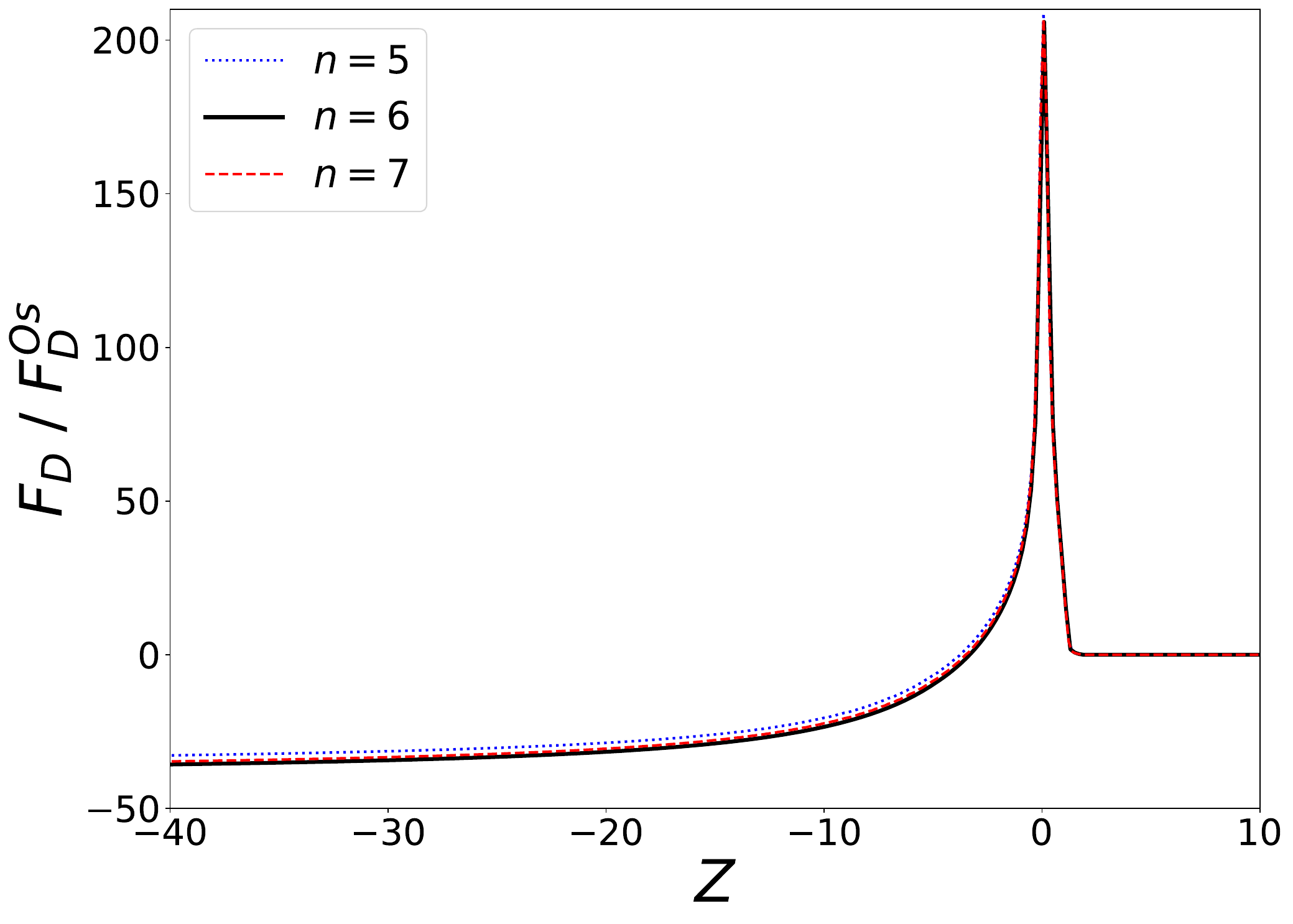}
    \caption{\textcolor{black}{$F_D^{\infty}/F_D^{Os}$ for for an isotropic wind with $u=5.0$ using different resolutions. Three resolution levels are shown ($n=5$, blue dotted line; $n=6$, solid black line; and $n=7$, red dashed line).}}
    \label{fig:resolution}
\end{figure}

\textcolor{black}{Figure~\ref{fig:r_in} shows $F_D^{\infty}/F_D^{Os}$ for isotropic and equatorial winds for different injection radii (in all cases $u=5$ is used). For the isotropic wind we obtain $F_D^{\infty}/F_D^{Os} = -40.82, -41.23, -42.14$. For the equatorial wind we obtain $F_D^{\infty}/F_D^{Os} = -67.21, -65.98, -63.14$. Thus, the difference is less than $\sim$3\% for the isotropic wind and $\sim$6\% for the equatorial wind for $r_0=0.20$, compared to using a smaller injection radius.
The difference in the value of the DF at $z\sim 0$ arises because its computation is over a different region for each case (as the integration in Equation~(\ref{eq:intDF}) starts at the location of the inner boundary). Therefore, the value of DF close to $z\sim 0$ increases for $r_0=0.10~R_0$ with respect to  $r_0=0.20~R_0$. Also, we note that the inner boundary is resolved with a different amount of cells for each case ($\sim$100 for $r_0=0.10~R_0$, $\sim$150 for $r_0=0.15~R_0$, and $\sim$200 for $r_0=0.20~R_0$).}
\begin{figure}
    \centering
    \includegraphics[width=\linewidth]{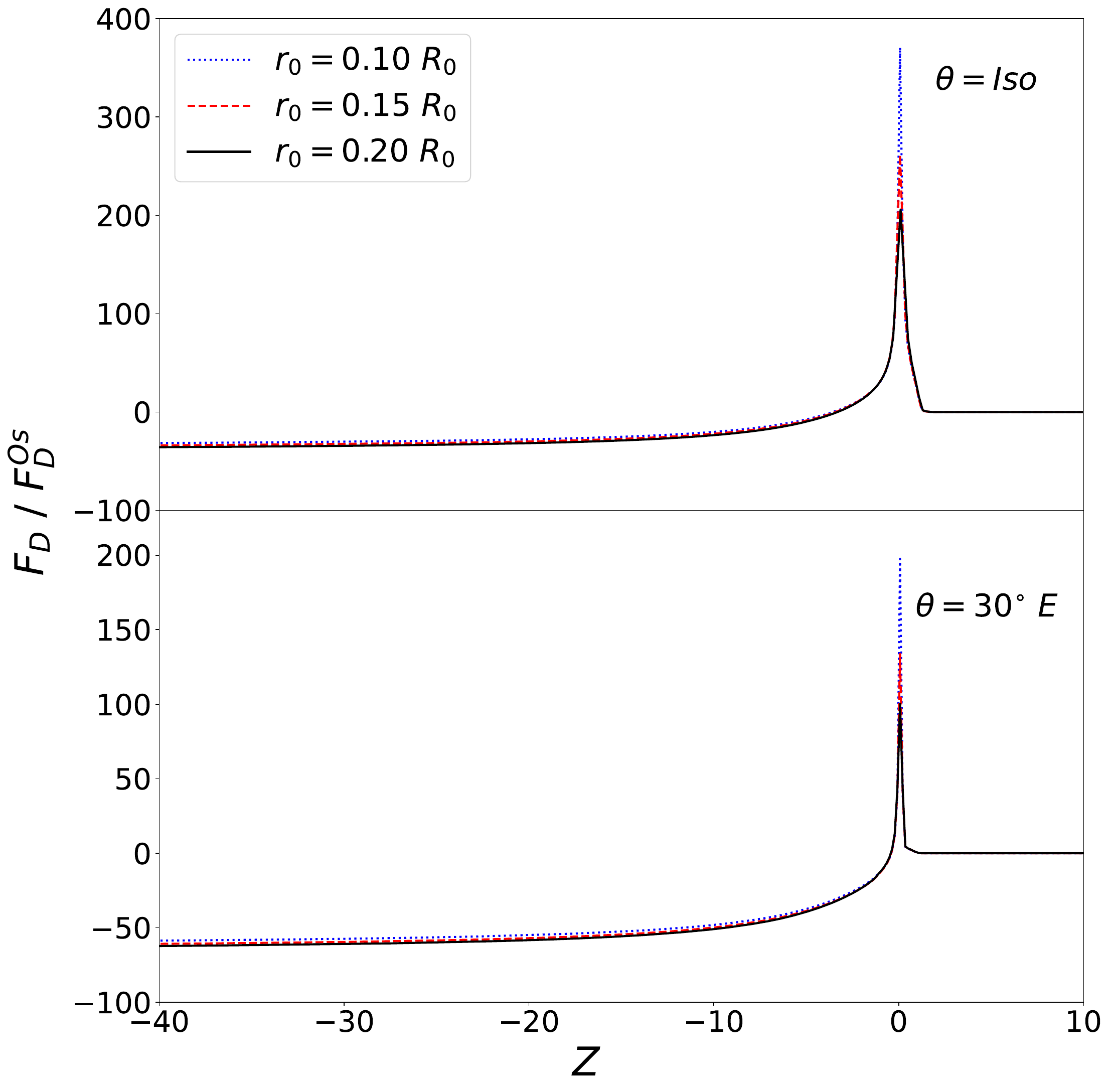}
    \caption{\textcolor{black}{$F_D^{\infty}/F_D^{Os}$ for an isotropic wind (upper panel) and equatorial wind with $\theta_w=30^{\circ}$ (lower panel) for different injection radii. For all cases, $u=5.0$. Three injection radius are shown ($r_0=0.10~R_0$, blue dotted line; $r_0=0.15~R_0$, red dashed line, $r_0=0.20~R_0$, solid black line).}}
    \label{fig:r_in}
\end{figure}

\section{Unit conversion}
\label{section:norm}
We use dimensionless units that can be rescaled to real physical units using the corresponding normalization factor (see Table~\ref{table:normalization}). The variables in cgs units are indicated by a
prime (e.g., the distance in cgs units is $r'$), and those in code units are indicated without a prime (e.g., the normalized distance is $r$).

The normalization factors for distance, density, velocity, and mass are $R_0'$, $\rho_a'$, $c_s'$, and $M'_{\star}$, respectively. This is $r' = r \cdot R_0'$, $\rho' = \rho \cdot \rho_a'$, $v' = v \cdot c_s'$, and $M' = M \cdot  M'_{\star}$. Since the Mach number is fixed at 5, the normalization factor for the sound speed is $c_s'=v_a'/M_a=v_a'/5$, thus $v' = v \cdot v_a'/5$. 

The normalization factor for the pressure is the ram pressure, hence $P' = P \cdot \rho_a' \ c_s'^2$. For the mass loss rate, we have $\dot{M}'= 4 \pi r'^2 \rho' v' = \dot{M} \cdot (R_0'^2 \ \rho_a' \ c_s')$. We set $R_0 = 1$, $\rho_{a} =1$, $c_s = 1$, and $M_a = 5$ in the code. For the time normalization factor, see Appendix~\ref{ap:critime}.

The conversion for the DF in code units and in physical units is:
\begin{equation}
    F_D'=F_D \cdot G \ M'_{\star} \ \rho_a' \ R_0',
\label{eq:intDF2}
\end{equation}
since the gravitational constant in the code is set to $G=1$. Meanwhile, for the case of $F_D^{Os}$, the conversion is:
\begin{equation}
    F_D^{Os\,'} = F_D^{Os} \cdot  \ G^2 \ M_{\star}'^2 \ \rho_a' \ c_s'^{-2}.
\label{eq:intDF3}
\end{equation}
Consequently, the normalization factor for the DF ratio is: $R_0' c_s'^2 / (G M'_{\star})$.

\section{Analytical models}
\label{section:appendix_soleq}
\subsection{DF equation}
\label{ap:solveeq5}
To compute analytically the DF, we consider a wind density profile with two components. Close to the star, the wind density scales as $\rho(r)=\rho_1 r_0^2/(R^2+z^2)$, where $z$ and $R$ are the vertical and radial cylindrical coordinates. Far from the star, the density is taken as constant inside a cylinder.
In the following, $r_0$ is the injection radius of the stellar wind and $r_1$ the final radius of the stellar wind, $z_0$ is the location of the bow-shock (along the $z$-axis), and $z$ is an arbitrary point away from the star. We consider only the DF produced by the stellar wind, neglecting the effect of the shocked ambient medium, which density differs from that of the environment at most by a factor of four. With these hypotheses, Equation~(\ref{eq:intDF}) reduces to:
\begin{eqnarray*}
F_D&=&4\pi GM_{\star}\int_{z_0}^{z}\int_{r_0}^{r_1} \frac{(\rho(r)-\rho_0) R dR z' dz'}{(R^2+z'^2)^{\frac{3}{2}}}    \\
&=&4\pi GM_{\star} \int_{z_0}^{z}\int_{r_0}^{r_1}\left[\frac{\rho_1 r_0^2}{(R^2+z^2)^{\frac{5}{2}}}-\frac{\rho_0}{(R^2+z^2)^{\frac{3}{2}}}\right] R dR z' dz'\\
&=&4\pi GM_{\star}\int_{z_0}^{z} \left[-\frac{\rho_1 r_0^2}{3}\left((r_1^2+z'^2)^{-\frac{3}{2}}-(r_0^2+z'^2)^{-\frac{3}{2}}\right) \right.\\
&&\left. -\rho_0\left((r_1^2+z'^2)^{-\frac{1}{2}}-(r_0^2+z'^2)^{-\frac{1}{2}}\right)\right] z'dz'=\\
&=&4\pi GM_{\star}\left[-\frac{\rho_1 r_0^2}{3}\left[(r_1^2+z^2)^{-\frac{1}{2}}-(r_0^2+z^2)^{-\frac{1}{2}}\right.\right.\\
&&\left.\left.-(r_1^2+z_0^2)^{-\frac{1}{2}}+(r_0^2+z_0^2)^{-\frac{1}{2}}\right]\right. -\rho_0\left[(r_1^2+z^2)^{\frac{1}{2}}\right.\\
&&\left.\left.-(r_0^2+z^2)^{\frac{1}{2}}.-(r_1^2+z_0^2)^{\frac{1}{2}}+(r_0^2+z_0^2)^{\frac{1}{2}}\right]\right],\\
\end{eqnarray*}
which can be re-written as:
\begin{eqnarray*}
    F_D(z)&=&A\left[(r_0^2+z^2)^{-\frac{1}{2}}-(r_1^2+z^2)^{-\frac{1}{2}}\right]\\
    &&+B\left[(r_0^2+z^2)^{\frac{1}{2}}-(r_1^2+z^2)^{\frac{1}{2}}\right]+F_D^{\infty},
\end{eqnarray*}
from where we obtain Equation~(\ref{eq:fdzinf}), this is:
\begin{eqnarray*}
F_D(z) &=& A \cdot H_1(z) + B \cdot H_2(z) +F_D^{\infty},
\end{eqnarray*}
where
\begin{eqnarray*}
    H_1(z) &=& (r_0^2+z^2)^{-\frac{1}{2}} - (r_1^2+z^2)^{-\frac{1}{2}}\;,\\
    H_2(z) &=& (r_0^2+z^2)^{\frac{1}{2}} - (r_1^2+z^2)^{\frac{1}{2}}\;,
\end{eqnarray*}
and the constants $A$, $B$, $F_D^{\infty}$ are:
\begin{eqnarray*}
    A&=&\frac{4\pi GM_{\star}\rho_1 r_0^2}{3}\;,\\ 
    B&=&4\pi\rho_0 GM_{\star}\;,\\
    F_D^{\infty}&=&A\left[(r_1^2+z_0^2)^{-\frac{1}{2}}-(r_0^2+z_0^2)^{-\frac{1}{2}}\right]\\
    &&+ B\left[(r_1^2+z_0^2)^{\frac{1}{2}}-(r_0^2+z_0^2)^{\frac{1}{2}}\right]\;. 
\end{eqnarray*}
The sign change of the constant $B$ shown in Table~\ref{tb:constdinf} happens because in the $u=5.0$ models, the shocked stellar wind has higher density than the ambient, while in the $u=0.2$ models the shocked stellar wind has lower density than the ambient, causing the quantity $\rho(r)-\rho_0$ to be positive or negative depending on the fitted model.

For $z\gg1$ and using a second-order Taylor approximation, we get:
\begin{eqnarray*}
    F_D(z)&\simeq& \frac{A}{2z}\left[r_1^2-r_0^2\right]+\frac{B}{2z}\left[r_0^2-r_1^2\right]+F_D^{\infty}\:.
\end{eqnarray*}
For the asymptotic case $z\to\infty$:
\begin{eqnarray*}
F_D(z \rightarrow \infty) &\equiv& F_D^{\infty}
\end{eqnarray*}
Figures~\ref{fig:ap1} and \ref{fig:ap2} show examples of fits obtained using this simple analytical description. The analytical model accurately reproduces both the behavior of the DF near the star, where the increase and drop in DF are due to the (approximately spherical) stellar wind, and at $z\ll 0$, where the DF is determined by the approximately cylindrical structure formed behind the star.

\begin{table}
	\centering
	\caption{Best-fit parameters $A$, $B$, $F_D^{\infty}$ and $r_i$ (each one normalized to $F_D^{Os}$) for the representative models shown in Figures~\ref{fig:ap1} and \ref{fig:ap2}. The $u$ parameter, $\theta_w$, and the wind orientation for each model are indicated. I stands for isotropic, P for polar, and E for equatorial.} 
	\label{tableA2}
	\begin{tabular}{|c|c|c|c|c|}
		\hline
		Model & $A/F^{Os}_D$ & $B/F^{Os}_D$ & $F^{\infty}_D/F^{Os}_D$ & $r_1$ \\
        ($u$, $\theta_w$, orientation) &   &              &                         &        \\ 
        \hline
        $(0.2, 90^{\circ},I)$  &  $0.30$ & $3.16$  & $7.33$  & $2.53$  \\
        $(5.0, 90^{\circ},I)$ & $40.36$ & $-3.58$ & $-40.82$ & $10.69$ \\
        $(5.0, 45^{\circ},P)$ & $85.09$  & $-0.47$  & $-18.41$  & $34.85$  \\
        $(0.2, 60^{\circ},E)$ & $0.78$  & $4.81$  & $9.89$  & $2.09$  \\
\hline
	\end{tabular}
    \label{tb:constdinf}
\end{table}

\begin{figure}
    \centering
    \includegraphics[width=\linewidth]{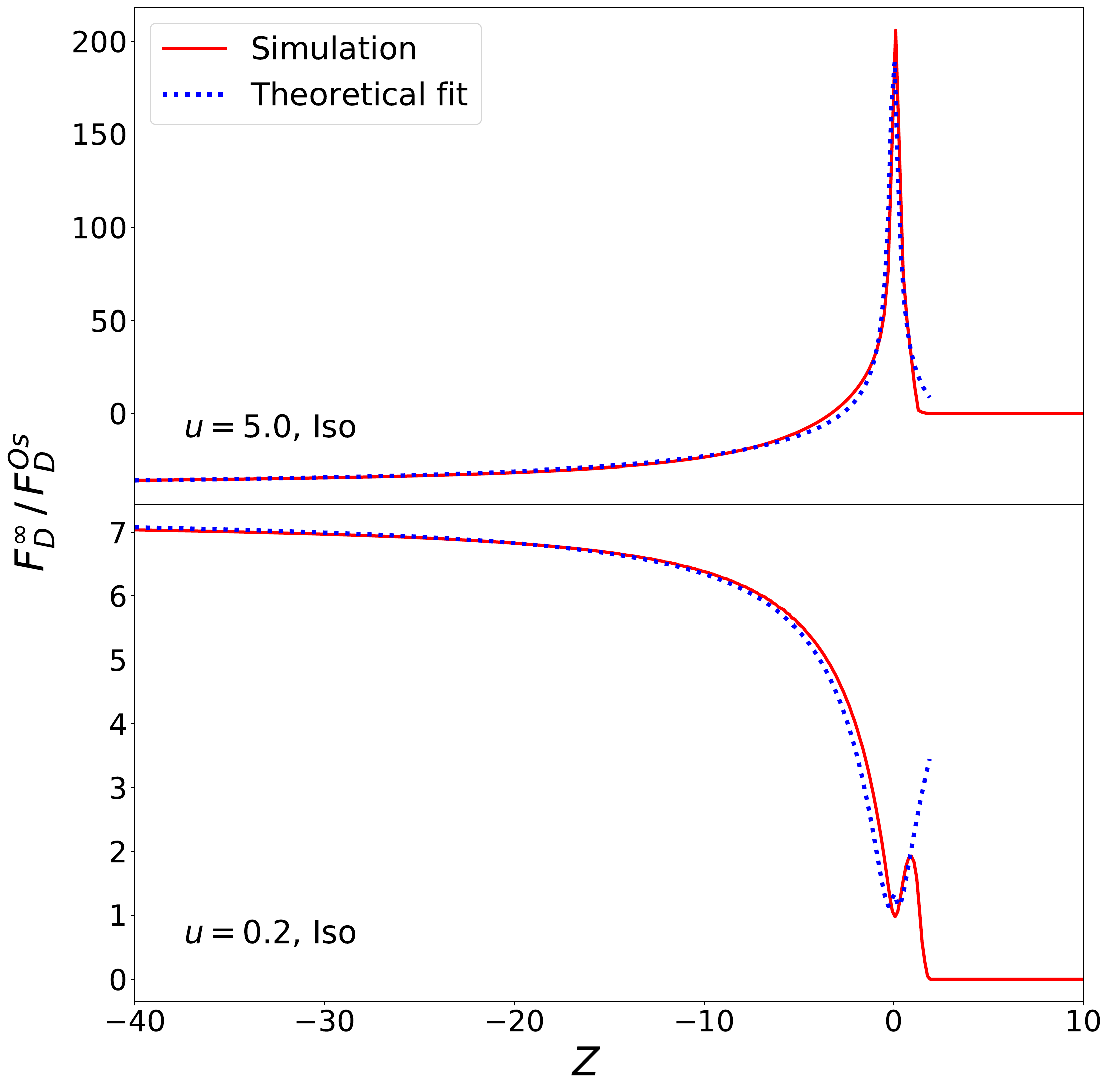}
    \caption{Comparison between the simulation (red solid line) and the fitted theoretical model (blue dotted line) for the isotropic case. The top panel shows the case for $u=5.0$ and the bottom panel for $u=0.2$.}
    \label{fig:ap1}
\end{figure}

\begin{figure}
    \centering
    \includegraphics[width=\linewidth]{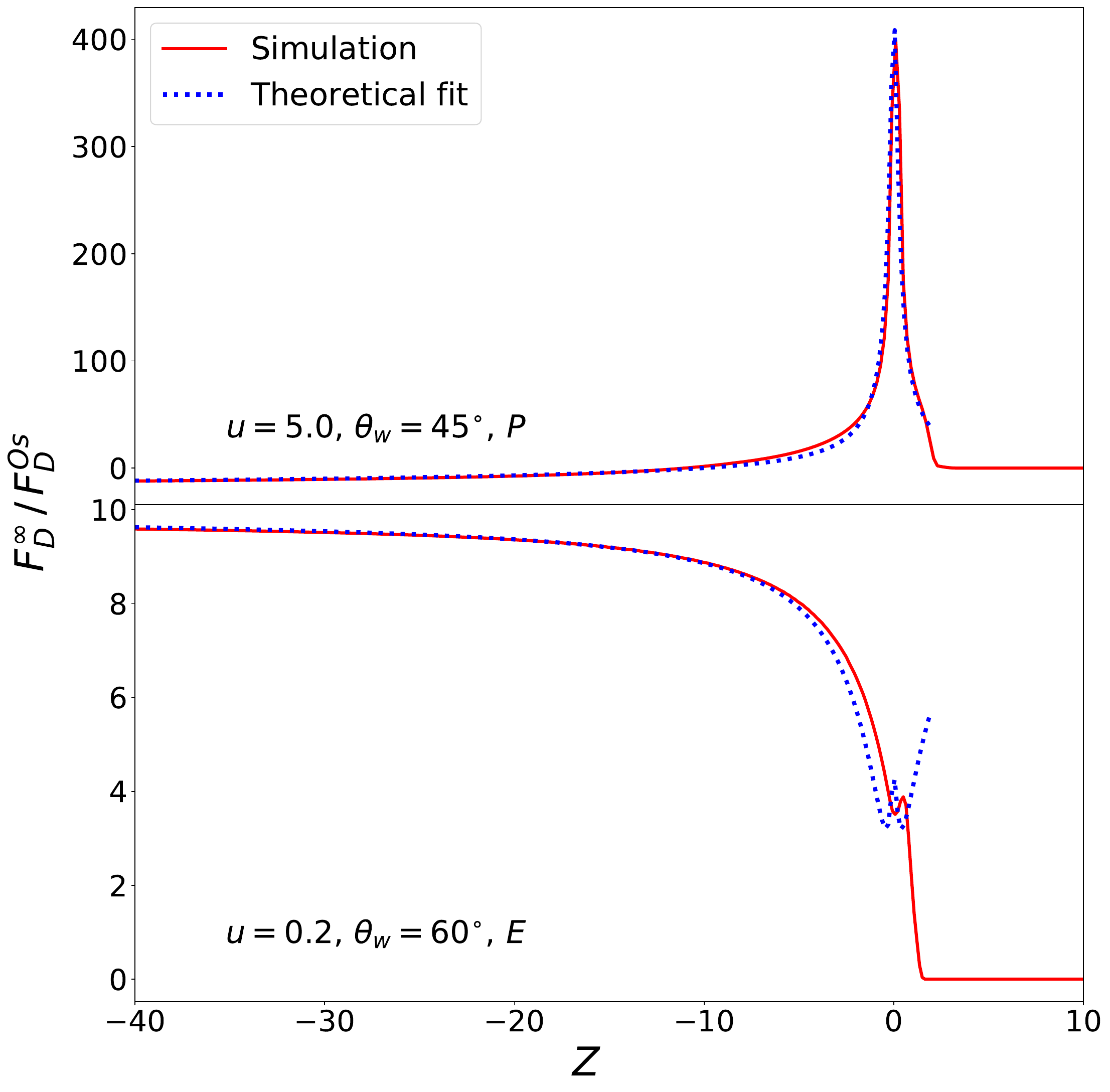}
    \caption{Comparison between the simulation (red solid line) and the fitted theoretical model (blue dotted line) for different winds. The top panel shows the case for a polar wind with $\theta_w=45^{\circ}$ and $u=5.0$. The bottom panel shows the case for an equatorial wind with $\theta_w=60^{\circ}$ and $u=0.2$.}
    \label{fig:ap2}
\end{figure}

\subsection{Differential equation}
\label{ap:solveeq6}
We solve Equation~(\ref{eq:DFsquare}) 
\begin{eqnarray*}
    \frac{du}{d\tau}&=&-P_0u^2-P_1u-P_2\;,
\end{eqnarray*}
by defining $u_m=-P_1/(2P_0)\approx -0.03$ as the value of $u$ corresponding to the maximum of the parabola, and $u_c$ as the physically plausible root of the equation, that is 
\begin{eqnarray*}
    u_c&=&u_m -\frac{\sqrt{P_1^2-4P_0P_2}}{2P_0}\approx 1.95\;,
\end{eqnarray*}
Then, we can rewrite the differential equation as
\begin{eqnarray}
\frac{du}{d\tau}&=&-P_0\left[\left(u+\frac{P_1}{2P_0}\right)^2-\left(\frac{P_1^2}{4P_0^2}-\frac{P_2}{P_0}\right)\right] \nonumber\\
&=&-P_0\left[\left(u-u_m\right)^2-\left(u_c-u_m\right)^2\right]\nonumber.
\end{eqnarray}
Thus, the time $\tau$ is:
\begin{eqnarray*}
    \tau = \int_0^{\tau}d\tau'&=&-\frac{1}{P_0}\int_{u_0}^{u}\frac{du'}{(u'-u_m)^2-(u_c-u_m)^2}\;.
\end{eqnarray*}
The integral admits two solutions, corresponding to the cases $u<u_0$ and $u>u_0$.\\
In the case $u_0<u_c$, we have:
\begin{eqnarray}
    \frac{\tau}{\tau_n}&=&\mbox{arctanh}\left(\frac{u-u_m}{u_c-u_m}\right)-\mbox{arctanh}\left(\frac{u_0-u_m}{u_c-u_m}\right)\;,
    \label{eq:tau}
\end{eqnarray}
where we have defined $\tau_n=1/[-P_0(u_c-u_m)]$.
For the limiting case $u_0=u_m$:
\begin{eqnarray*}
    \frac{\tau}{\tau_n}&=&\mbox{arctanh}\left(\frac{u-u_m}{u_c-u_m}\right)\;,
\end{eqnarray*}
thus we obtain:
\begin{eqnarray*}
    u&=&u_m+(u_c-u_m)\,\mbox{tanh}\left(\frac{\tau}{\tau_n}\right)
\end{eqnarray*}
In the case $u_0>u_c$:
\begin{eqnarray*}
    \frac{\tau}{\tau_n}&=&\mbox{arccoth}\left(\frac{u-u_m}{u_c-u_m}\right)-\mbox{arccoth}\left(\frac{u_0-u_m}{u_c-u_m}\right)\;,
\end{eqnarray*}
For the limiting case $u_0\to\infty$:
\begin{eqnarray*}
    \frac{\tau}{\tau_n}&=&\mbox{arccoth}\left(\frac{u-u_m}{u_c-u_m}\right)\;,
\end{eqnarray*}
thus we obtain:
\begin{eqnarray*}
    u&=&u_m+(u_c-u_m)\,\mbox{coth}\left(\frac{\tau}{\tau_n}\right)\;.
\end{eqnarray*}

\subsection{Time normalization in code units}
\label{ap:taun}
The DF acceleration ($du / dt$) is:
\begin{eqnarray*}
    \frac{du}{d t} &=& -\frac{F_D^{\infty}}{M_{\star} v_w}\;,
\end{eqnarray*}
where $F_D^{\infty}$ is the DF at infinity, $M_{\star}$ is the mass of the star, and $v_w$ is the stellar wind velocity.\\
Normalizing by $F_D^{Os}$ (which is a constant), one obtains:
\begin{eqnarray*}
    \frac{du} {d\left(t \ F_D^{Os}\right)} &=& -\frac{F_D^{\infty}}{F_D^{Os}}\frac{1}{M_{\star} v_w}\;.
\end{eqnarray*}
Rearranging terms and assuming that the mass and stellar wind do not change much, then:
\begin{eqnarray*}
    \frac{du} {d\left(t \ F_D^{Os} \ / (M_{\star} v_w)\right)} &=& -\frac{F_D^{\infty}}{F_D^{Os}}\;.
\end{eqnarray*}
Thus, the dimensionless time parameter is $\tau = t \ F_D^{Os} \ / (M_{\star} v_w)$ with which we obtain Equation~(\ref{eq:DFsquare}). 

\subsection{Critical time}
\label{ap:critime}
Variables with the prime (') symbol represent quantities in cgs units as explained in Appendix~\ref{section:norm}. The law of motion for the velocity of the star, $v_a$, is:
\begin{eqnarray*}
    \frac{dv_a'}{dt'}&=&-\frac{F_D^{\infty\,'}}{M'_{\star}}\;.
\end{eqnarray*}
Since $u'=u$ and dividing by $v_w$ (assuming that the wind velocity is constant in time) and by $F_D^{Os\,'}$ (which is also taken as constant), then:
 \begin{eqnarray*}
     \frac{M'_{\star}v_w'}{F^{Os\,'}_{D}}\frac{du}{dt'}&=&-\frac{F_D^{\infty\,'}}{F^{Os\,'}_{D} }\;.
 \end{eqnarray*}
Using the conversion factor for the DF ratio shown in Table~\ref{table:normalization} (and derived in Appendix~\ref{section:norm}), and rearranging terms, we have:
\begin{eqnarray*}
    \frac{du}{d\left(t'\frac{R_0'c_s'^2}{GM'_{\star}}\frac{F^{Os\,'}_{D}}{M'_{\star}v_w'}\right)} &=& \frac{du}{d\tau} = -\frac{F_D^{\infty}}{F^{Os}_{D}}\;,
\end{eqnarray*}
where $\tau=t'\frac{R'_0c_s'^2F^{Os\,'}_{D}}{GM_{\star}'^2v_w'}$ is the dimensionless time parameter.\\
Thus, using Equation~(\ref{eq:ostriker}) with $\Lambda = 125$, $M_a=v_a'/c_s'= 5$, $c_s' = v_a'/5$, and replacing $R_0$ as defined in W96, we get:
\begin{eqnarray*}
    t' &=& \tau \frac{25 v_w'^{1/2} v'_a} {\sqrt{4\pi} G \dot{M}^{'\,1/2} \rho_a^{'\,1/2} \ln\left[ 25 \times 24^{1/2} \right]}\;.
\end{eqnarray*}
Finally, using $\tau_c\approx0.2\tau_n=0.05$ and the number density, the critical time in cgs units is:
\begin{eqnarray*}
    t'_c &=& 0.07\,G^{-1}\ m_H^{-1/2} \ n_a'^{-1/2} \ \dot{M}'^{-1/2} \ v_w'^{1/2} \ v'_a\;.
\end{eqnarray*}
Using typical values for the ambient medium and stellar wind, we recover Equation~(\ref{tauncgs}).

\subsection{Equation of movement}
\label{ap:rnorm}
The star velocity is:
\begin{eqnarray*}
    \frac{dx}{d t} &=& u v_w\;,
\end{eqnarray*}
where $x$ is the displacement of the star, $v_w$ is the stellar wind velocity, and $u=v_a/v_w$ is the velocity of the star to wind ratio. Replacing the time by the dimensionless time parameter $\tau = t \ F_D^{Os} \ / (M_{\star} v_w)$, we get
\begin{eqnarray*}
    \frac{F_D^{Os}}{M_{\star}v_w^2}\frac{dx}{d \tau} &=& u\;.
\end{eqnarray*}
Rearranging terms and assuming that the stellar mass and wind are constant in time, we get:
\begin{eqnarray*}
    \frac{d\left(x \ F_D^{Os} \ / (M_{\star} v_w^2)\right)}{d\tau}  &=& u\;.
\end{eqnarray*}
Thus, the dimensionless displacement parameter is $r = x \ F_D^{Os} \ / (M_{\star} v_w^2)$. For the parameters in cgs units we get:
\begin{eqnarray*}
    r & = & 60.42\frac{G^2M_{\star}\rho_a}{v_a^2v_w^2}x\\
      & = & 2.76\times10^{-13}\left(\frac{M_{\star}}{M_{\odot}}\right)\left(\frac{n_a}{{\rm cm}^{-3}}\right)\\
      &&\times\left(\frac{v_a}{{\rm 100\,km\,s}^{-1}}\right)^{-2}\left(\frac{v_w}{{\rm 100\,km\,s}^{-1}}\right)^{-2}\left(\frac{x}{\rm pc}\right)\;. 
\end{eqnarray*}
For $n_a=10^4$~cm$^{-3}$, $x=20$~pc, and the initial stellar velocity $v_a=10$~km~s$^{-1}$  we have:
\begin{eqnarray*}
    r & = & 5.52\times10^{-6}\left(\frac{M_{\star}}{M_{\odot}}\right)\left(\frac{v_w}{{\rm 100\,km\,s}^{-1}}\right)^{-2}\;.
\end{eqnarray*}
Thus, for  $M_{\star}=40~M_{\odot}$ and $v_w=1000$~km~s$^{-3}$ (i.e., $u_0=0.01$ for a Wolf-Rayet star), we get $r=2.21\times10^{-6}$.

Integrating $dr/d\tau = u$, and considering Equation~(\ref{eq:tau}) from Appendix~\ref{ap:solveeq6}, we get:
\begin{eqnarray*}
r=\int_0^rdr'&=&\int_0^{\tau}\left[u_m+(u_c-u_m)\mbox{tanh}\left( \frac{\tau'}{\tau_n}+C_0\right)\right]d\tau'\\
    &=&u_m\tau+(u_c-u_m)\tau_n\ln\left[\mbox{cosh}\left(\frac{\tau}{\tau_n}+C_0\right)\mbox{sech}\left(C_0\right)\right]\;,
\end{eqnarray*}
where $C_0=\mbox{arctanh}[(u_0-u_m)/(u_c-u_m)]$.

Solving for $\tau_n=0.25$, $u_c=1.95$, $u_m=-0.03$, $r\approx2.21\times10^{-6}$  and $C_0 = 0.02$ we get $\tau\approx2.71\times10^{-4}$. 

Then the final velocity ($v_a$) for each $\tau$ is:
\begin{eqnarray*}
    v_a & = & v_w\left[u_m+(u_c-u_m)\mbox{tanh}\left( \frac{\tau}{\tau_n}+C_0\right)\right]\;,
\end{eqnarray*}
where we have considered $v_a=v_wu$. Thus, for the 
Wolf-Rayet star (with $\tau\approx2.71\times10^{-4}$) we get $v_a\approx 11.99~$km~s$^{-1}$.

\bsp	
\label{lastpage}
\end{document}